\documentclass[aps,a4paper,pre,twocolumn,showpacs,floatfix]{revtex4}

\usepackage{graphicx}
\usepackage{amsmath}
\newcommand{\mypic}{8cm}
\newcommand{\et}{{\it et al.}\ }
\newcommand{\coex}{{\rm \ast}}

\begin{document}

\title{Phase behavior and structure of model colloid-polymer mixtures confined between two parallel planar walls}

\author{Andrea Fortini$^1$\footnote{Electronic address: a.fortini@phys.uu.nl}, Matthias Schmidt$^{2}$\footnote{This work
   was initiated while at Soft Condensed Matter, Debye Institute,
   Utrecht University, Princetonplein 5, 3584 CC Utrecht, The
   Netherlands.} and Marjolein Dijkstra$^1$}

\affiliation{$^1$ Soft Condensed Matter, Debye Institute, Utrecht
University, Princetonplein 5, 3584 CC Utrecht, The Netherlands.}
\affiliation{$^2$ H.H. Wills Physics Laboratory, University of
Bristol, Tyndall Avenue, Bristol BS8 1TL, United Kingdom}
\affiliation{Institut f\"ur Theoretische
   Physik II, Heinrich-Heine-Universit\"at D\"usseldorf,
   Universit\"atsstra\ss e 1, D-40225 D\"usseldorf, Germany.}


\begin{abstract}
Using Gibbs ensemble Monte Carlo simulations and density
functional theory we investigate the fluid-fluid demixing
transition in inhomogeneous colloid-polymer mixtures confined
between two parallel plates with separation distances between one
and ten colloid diameters covering the complete range from quasi
two-dimensional to bulk-like behavior. We use the
Asakura-Oosawa-Vrij model in which colloid-colloid and
colloid-polymer interactions are hard-sphere like, whilst the pair
potential between polymers vanishes.  Two different types of
confinement induced by a pair of parallel walls are considered,
namely either through two hard walls or through two semi-permeable
walls that repel colloids but allow polymers to freely penetrate.
For hard (semi-permeable) walls we find that the capillary binodal
is shifted towards higher (lower) polymer fugacities and lower
(higher) colloid fugacities as compared to the bulk binodal; this
implies capillary condensation (evaporation) of the colloidal
liquid phase in the slit. A macroscopic treatment is provided by a
novel symmetric Kelvin equation for general binary mixtures, based
on the proximity in chemical potentials of statepoints at capillary
coexistence and the reference bulk coexistence. Results for
capillary binodals compare well with those obtained from the classic
version of the Kelvin equation due to Evans and Marini Bettolo
Marconi [J. Chem. Phys. {\bf 86}, 7138 (1987)], and are
quantitatively accurate away from the fluid-fluid critical point,
even at small wall separations. However, the significant shift of
the critical polymer fugacity towards higher values upon
increasing confinement, as found in simulations, is not
reproduced. For hard walls the density profiles of polymers and
colloids inside the slit display oscillations due to packing
effects for all statepoints. For semi-permeable walls either
similar structuring or flat profiles are found, depending on the
statepoint considered.
\end{abstract}

\maketitle

\section{Introduction}
Capillary effects that are induced by the confinement of a system
are crucial to a variety of phenomena. An everyday example is the
capillary rise, against gravity, of the meniscus of a free
gas-liquid interface at the wall of a container that encompasses
the fluid. The contact angle at which the gas-liquid interface
hits the wall is described by Young's equation \cite{JRBW82},
$\gamma_{\text{lg}} \cos
\theta=\gamma_{\text{wg}}-\gamma_{\text{wl}}$, where the relevant
interfacial tensions are those between the coexisting liquid and
gas phase, $\gamma_{\text{lg}}$, the wall and the gas phase,
$\gamma_{\text{wg}}$, and the wall and the liquid phase,
$\gamma_{\text{wl}}$. When the contact angle is finite, $0 <
\theta < \pi$, the liquid phase partially wets the wall; for
$\theta=0$ a macroscopic layer of liquid grows between the wall
and the gas phase far away from the wall, hence the liquid
completely wets the wall; for $\theta=\pi$ a correspondingly
reversed situation occurs: a macroscopic layer of gas grows
between the wall and the liquid phase far away from the wall,
hence the wall is completely ``wet'' by the gas phase and drying
occurs. All these phenomena are driven by the influence of a
{\em single} wall on the fluid.

Different, but related effects occur under confinement of fluids
in pores, e.g.\ between {\em two} parallel planar walls. The phase
diagram of such a confined system can differ significantly from
that in bulk \cite{Gelb1999,RE90}. Depending on the nature of the
interactions between fluid particles and the walls, the bulk gas
with chemical potential $\mu<\mu_\text{sat}$, where
$\mu_\text{sat}$ is the value at saturation, can condense inside
the pore and form a dense confined liquid phase. This capillary
phase transition is accompanied by a jump in the adsorption
isotherm at a given value of $\mu<\mu_\text{sat}$.  Confinement
may lead to stabilization of a phase that is unstable (or at least
metastable) in bulk for a given statepoint. As opposed to
capillary condensation the opposite effect, referred to as
capillary evaporation, is also feasible: A fluid with chemical
potential $\mu > \mu_\text{sat}$ forms a liquid in bulk, but
evaporates inside the capillary. A simple, yet powerful, way to
quantitatively describe capillary phase transitions is based on
the Kelvin equation, derived from a macroscopic treatment of the
thermodynamics of the inhomogeneous system. For capillaries with
planar slit-like geometries \cite{RE87}, it predicts that
liquid-gas coexistence inside the slit will occur at $\mu =
\mu_\text{sat} + \gamma_\text{lg} \cos(\theta)/h$, where $h$ is
the separation distance between the two walls. Hence, contact
angles $0 \leq \theta < \pi/2$ correspond to capillary
condensation, while $\pi/2 < \theta \leq \pi$ corresponds to
capillary evaporation.

In this work we investigate capillary phenomena occurring on a
mesoscopic scale using a simple model for a mixture of
sterically-stabilized colloidal particles and non-adsorbing
polymer coils confined between two parallel planar walls. 
Using colloids as model systems offers advantages over molecular
substances due to easy experimental access to the relevant length and time scales, and the possiblilty of using real space techniques like
confocal microscopy.  Our work is intended to stimulate such
investigations in order to gain further insights into capillary
phenomena. The topology of the bulk phase diagram of colloid-polymer mixtures depends on the polymer size (as compared to the size of the
colloids) and the polymer concentration. At sufficiently high polymer concentration and for sufficiently large polymer, the bulk system demixes into a
{\em colloidal liquid} phase that is rich in colloids (and dilute
in polymers), and a {\em colloidal gas} phase that is dilute in
colloids (and rich in polymers)
\cite{EJM91,EJM94,SMI95,MDJMB99,MD01}. This phenomenology is
similar to gas-liquid coexistence in simple fluids with the
polymer fugacity $z_\text{p}$ playing the role of inverse
temperature, and hence governing the strength of interparticle
attractions. However the phase separation in colloid-polymer
mixtures is purely entropy-driven, due to an effective depletion
interaction between the colloidal particles, that is mediated by
the polymer \cite{WCKP02,RT03}.  A similar mechanism induces an
effective depletion interaction between a colloidal particle and a
planar hard wall \cite{JMB01}. For a review on the properties of
colloid-polymer mixtures in contact with a hard wall, we refer the
reader to Brader \et \cite{JMB03}. Recently the wall-fluid
interfacial tensions \cite{DGALA04a,PPFW04a,Fortini2005a} and the
contact angle of the free interface and a hard wall \cite{PPFW04},
and the gas-liquid interfacial tension \cite{RLCV04c,RLCV04} have
been studied theoretically and with computer simulations. Moreover
complete wetting of the colloidal liquid phase at a hard wall is
found, close to the critical point, in both simulation
\cite{MDRvR02,MDRvR05,MDRvRRRAF05} and theory \cite{JMB03,JMB02}.
Complete wetting of the colloidal liquid is also observed in
experimental realizations of colloid-polymer mixtures in contact
with glass substrates \cite{DGALA03,DGALA04} or glass substrates
that possess the same coating with an organophilic group as the
colloidal particles do \cite{WKW03}.  On the other hand,
experiments \cite{WKW03-a} with polymer-grafted substrates (of the
same chemical nature as the dissolved polymers) showed that the
contact angle is larger than $\pi/2$.  Although the structure of
the polymer coating was not studied in detail, Wijting \et
\cite{WKW03-a} expect the polymers to form a fluffy layer with a
distribution of loops and tails. Such a layer is known to diminish
the depletion interaction between colloidal particles and the
substrate \cite{PJ96}. Wessels \et \cite{PPFW04a} modeled this
type of substrate using a semi-permeable wall, that is completely
penetrable to the polymers but act like a hard wall on the
colloids. Using DFT they predict complete drying. More recently
the behavior of the mixtures in contact with porous walls was
investigated, and both wetting of the surface and drying into the
porous matrix depending on the precise path in the phase diagram
were found \cite{Wessels2005}. Porous walls are wet by the liquid
phase, with a transition from partial to complete wetting at a
polymer fugacity $z_\text{p}$ almost independent on the porosity
of the wall.

Although much research has been devoted to the understanding of
the behavior of the mixture in contact with a single wall, few
studies have been reported on the effect of confinement on the
phase behavior of the mixture.  Computer simulations and theory
were used to study porous matrices. When impenetrable to both
components they were found to induce capillary condensation,
whereas matrices penetrable to the polymers, but not to the
colloids, induce capillary evaporation \cite{MSES03,PPFW03}.
Furthermore laser induced confinement\ \cite{IOG+03} was found to
induce capillary condensation. The only experimental result on
capillary condensation that we are aware of is that by Aarts \et
\cite{DGALA04}.  These authors found capillary condensation for
colloid-polymer mixtures confined in a wedge formed by a glass
bead placed on a glass substrate.

The aim of the present paper is to determine the phase behavior
and the structure of colloid-polymer mixtures confined between two
parallel planar walls for a complete range of wall separation
distances. We focus on the fluid part of the phase diagram, and
present results for hard walls and semi-permeable walls, obtained
from computer simulations and density functional theory.  A
preliminary account of this work was given in Refs.\
\cite{MSAF03,MSAF04}. Here we supply all relevant details of the
simulation method. We also investigate the structure of the
mixtures inside the pore, and derive and test a generalized Kelvin
equation for binary mixtures confined in slit-like pores and
compare its predictions with our simulation results.

\begin{figure}[htbp]
  \begin{center}
    \includegraphics[width=\mypic]{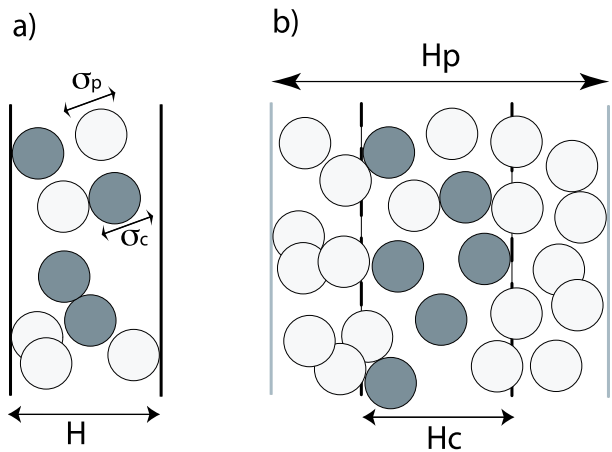}
    \caption{ (a) Illustration of the AOV model of hard sphere
      colloids (gray circles) of diameter $\sigma_\text{c}$ and ideal
      polymers (transparent circles) of diameter $\sigma_\text{p}$
      confined between parallel hard walls of area $A$ and separation
      distance $H$.  Colloids behave as hard spheres, polymers cannot
      penetrate colloids, and polymers may freely overlap. The walls
      are impenetrable to both components. The $z$-axis is
      perpendicular to both walls, and the origin is located in the
      middle of the slit. (b) Same as in (a) but for walls that are
      penetrable for polymers and impenetrable for colloids
      (semi-permeable walls) of area $A$ and separation distance
      $H_\text{c}$.  Polymers are confined between parallel hard walls
      (continuous line) with separation distance $H_\text{p}$.  This
      is a model for substrates at distance $H_\text{p}$ coated with a
      polymer brush (not shown) of thickness
      $(H_\text{p}-H_\text{c})/2$; the free distance between both
      polymer brushes is $H_\text{c}$.  Solute polymers (open spheres)
      are able to penetrate the brush (dashed lines) but not the
      substrate; the brush acts like a hard wall to the colloids.}
    \label{F:model}
  \end{center}
\end{figure}

\section{Model}
\label{mod} Consider a binary mixture of $N_\text{c}$ hard spheres
of diameter $\sigma_\text{c}$ representing the colloidal particles
and $N_\text{p}$ non-interacting spheres of diameter
$\sigma_\text{p}$ representing the (ideal) polymers in a volume
$V$ at temperature $T$.  This is a simple model for a
colloid-polymer mixture as the interaction between
sterically-stabilized colloids can be made such that it resembles
closely that of hard spheres, and a dilute solution of polymers in
a $\theta$-solvent is weakly interacting. Colloids and polymers
interact via a hard-sphere-like potential, as the polymers are
excluded from a centre-of-mass distance
$(\sigma_\text{c}+\sigma_\text{p})/2$ from the colloid, where
$\sigma_\text{p}=2R_g$, and $R_g$ is the radius of gyration of the
polymer coils.  We treat the solvent as an inert continuum.  This
so-called Asakura-Oosawa-Vrij (AOV) model \cite{SAFO54,Asakura1958,AV76} is
defined by the pair potentials
\begin{equation}
v_\text{cc}(R_{ij})= \left \{ \begin{array}{ll}
\infty & \textrm{ if $R_{ij} < \sigma_\text{c}$ } \\
0 & \textrm{otherwise},
\end{array} \right.
\end{equation}
where $R_{ij}=|\vec{R}_i-\vec{R}_j|$ is the distance between two
colloidal particles, with $\vec{R}_i$ the position of the
centre-of-mass of colloid $i$,
\begin{equation}
v_\text{pp}(r_{ij})= 0  ,
\end{equation}
where $r_{ij}=|\vec{r_i}-\vec{r_j}|$ is the distance between two
polymers, with $\vec{r_i}$ the position of the centre-of-mass of
polymer $i$ and
\begin{equation}
v_\text{cp}(|\vec{R}_i-\vec{r}_j|)= \left \{ \begin{array}{ll}
\infty & \textrm{ if $|\vec{R}_i-\vec{r}_j| < (\sigma_\text{c}+\sigma_\text{p})/2$ } \\
0 & \textrm{ otherwise,}
\end{array} \right.
\end{equation}
where $|\vec{R}_i-\vec{r}_j|$ is the distance between colloid $i$ and
polymer $j$.

The size ratio $q=\sigma_\text{p}/\sigma_\text{c}$ is a geometric
parameter that controls the range of the effective depletion
interaction between the colloids. We denote the packing fraction
by $\eta_k=(\pi\sigma_k^3 N_k)/(6 V)$, with $k=\text{c, p}$ for
colloids and polymers, respectively. As alternatives to
$\eta_\text{p}$, we use as a thermodynamic variable the colloidal
fugacity $z_\text{c}$, or the polymer reservoir packing fraction
$\eta_\text{p}^\text{r}$ that satisfies the (ideal gas) relation
\begin{equation}
  \eta_\text{p}^\text{r} =
  \frac{\pi}{6} \sigma_\text{p}^3 z_\text{p}. \label{pacres}
\end{equation}

The hard walls are modeled such that neither polymers nor colloids can
penetrate the walls.  The wall-particle potential acting on species
$k=\text{c, p}$ is
\begin{equation}
  v_{\text{w}k}(z)= \left \{ \begin{array}{ll}
    0 & \textrm{if $  -(H - \sigma_k)/2  < z  < (H-\sigma_k)/2 $ } \\
    \infty & \textrm{otherwise,}
  \end{array} \right.
\end{equation}
where $z$ is the coordinate normal to the walls, and $H$ is the wall
separation distance. We define the volume of the system as $V=AH$,
where $A$ is the (lateral) area of the confining walls. Fig.\
\ref{F:model}(a) shows an illustration of the model.

Semi-permeable walls are defined by the external potential
\begin{equation}
  v_{\text{w}k}(z)= \left \{ \begin{array}{ll}
    0 & \textrm{if $-(H_k-\sigma_k)/2  < z  < (H_k-\sigma_k)/2 $ } \\
    \infty & \textrm{otherwise,}
  \end{array} \right.
\end{equation}
where $H_\text{c}$ is the wall separation distance that the colloids
experience, while $H_\text{p}$ is the wall separation distance that
the polymers feel; the latter can be interpreted as the distance
between substrate walls.  We define the volume as
$V=AH_\text{c}$. Fig.\ \ref{F:model}(b) shows an illustration of this
model.  Any choice of $H_\text{p} > H_\text{c} +2 \sigma_\text{p} $
leads to decoupling of the (inner) colloid and (outer) polymer wall,
since the ideal gas of polymers cannot mediate correlations from the
substrate to the interior of the system.

\section{Simulation method}
\label{sim} We perform Gibbs Ensemble Monte Carlo (GEMC)
simulations to determine the phase behavior of the AOV model in bulk
and confined between two parallel planar walls. The number of
colloids $N_\text{c}$, the number of polymers $N_\text{p}$, and the
volume $V$ are kept fixed and are divided into two separate subsystems
a and b with volume $V^\text{a}$ and $V^\text{b}$, respectively, with
the constraints that $V=V^\text{a}+V^\text{b}$ ,
$N_\text{c}=N_\text{c}^\text{a}+N_\text{c}^\text{b}$ and
$N_\text{p}=N_\text{p}^\text{a}+N_\text{p}^\text{b}$. The two
subsystems are allowed to exchange both particles and volume in order
to satisfy the conditions for phase equilibrium, i.e., equality in
both phases of the chemical potentials of the two species and of the
pressure. The method can be applied to bulk as well as to confined
systems \cite{AZP87,AZP87b}, and will be described briefly below; for
more details, we refer the reader to Ref.\ \cite{DFBS96}.

In more detail, the bulk equilibrium conditions between two coexisting
phases require equal temperature $T$, equal chemical potential $\mu_i$
for each species $i$ ($i=\text{c, p}$ in our case) and equal pressure
$P$.  In our model, the temperature $T$ is irrelevant; because of the
hard-core nature of the interaction potentials it acts only as a
scaling factor setting the thermal energy scale.  In the GEMC method,
the two coexisting phases are simulated simultaneously in two separate
(cubic) boxes with standard periodic boundary conditions. The
acceptance probability for a trial move to displace a randomly
selected particle is
\begin{equation}
  \mathcal{P}=\text{min}\{1,\exp[ -\beta( U_\text{new}-U_\text{old}) ] \},
  \label{p1}
\end{equation}
where
\begin{equation}
  \exp [ -\beta ( U_\text{new}-U_\text{old}) ]= \left \{ \begin{array}{ll}
    1 & \textrm{ if $\beta U_\text{new}=0$ } \\
    0 & \textrm{ if $\beta U_\text{new}=\infty $  }
\end{array} \right.
\label{side}
\end{equation}
with $U_\text{new}$ and $U_\text{old}$ the energy of the new and old
generated configuration, and $\beta=1/(k_B T)$ the inverse temperature
with $k_B$ being the Boltzmann constant.  Note that $U_\text{old}=0$
as configurations with $U_\text{old}=\infty$ are excluded by a
vanishing Boltzmann factor (regardless of the temperature).  Transfer
of (single) particles between the two boxes is used to satisfy equal
chemical potential for both species. We select at random which
subsystem is the donor and which is the recipient as well as the
species (colloid or polymer) of the particle transfer. Subsequently, a
specific particle is randomly selected in the donor box and
transferred to a random position in the recipient box with probability
\begin{equation}
  \mathcal{P}=\text{min}\left\{1,\frac{N_i^\text{a}
    V^\text{b}}{(N_i^\text{b}+1)V^\text{a}} \exp[ -\beta (
    U_\text{new}-U_\text{old})]\right\}, \label{p2}
\end{equation}
where $i=\text{c, p}$ is the species of the selected particle,
$V^\text{a}$ is the volume of the donor box and $V^\text{b}$ is the
volume of the recipient box.

In addition, volume changes of the boxes are used to satisfy the
condition of equal pressure in both subsystems. We hence perform a
random walk in $\ln(V_a/V_b)$ with the acceptance probability
\begin{equation}
  \mathcal{P}=\text{min} \left \{1,\mathcal{R}
  \exp[ -\beta ( U_\text{new}-U_\text{old})]  \right\},
\label{p3}
\end{equation}
where
\begin{equation}
  \mathcal{R}= \left( \frac{V^\text{a}_\text{new}}{V^\text{a}_\text{old}}
  \right)^{N_\text{c}^\text{a}+N_\text{p}^\text{a}+1}
  \left( \frac{V^\text{b}_\text{new}}{V^\text{b}_\text{old}}
  \right)^{N_\text{c}^\text{b}+N_\text{p}^\text{b}+1},
\end{equation}
and with the condition that the total volume
$V=V^\text{a}_\text{old}+V^\text{b}_\text{old}=V^\text{a}_\text{new}+V^\text{b}_\text{new}$
is constant , where the subscript $\text{old}$ ($\text{new}$)
marks quantities before (after) the trial displacement.

Phase equilibria in confined geometries, like e.g.\ inside slit pores
as considered here, can be determined by either determining the
adsorption isotherm in simulation or by employing the GEMC simulation
method extended to a slit pore \cite{AZP87b}.  In the first method,
the pore is coupled to a reservoir of bulk fluid. The adsorption of
the fluid inside the pore is then measured at constant temperature for
different bulk densities of the reservoir, i.e.\ the chemical
potentials of the various species are fixed. A jump in the adsorption
isotherm corresponds to capillary condensation or evaporation in the slit pore.
However, this method is inaccurate for determining phase coexistence
due to hysteresis.

In order to determine the binodal lines, we hence employ the GEMC
method adapted to a slit pore \cite{AZP87b}: Two separate
simulation boxes are simulated, one containing the confined gas
and one containing the confined liquid.  Each box has periodic
boundary conditions in both directions parallel to the walls.
Capillary coexistence implies equality of temperature, chemical
potentials for both species as well as equality of the wall-fluid
interfacial tension. Fulfilling the first two conditions is
performed similar to the procedure for determining bulk phase
equilibria, i.e.\ using particle displacements with acceptance
probability given by Eq.\ (\ref{p1}) and particle transfers with
acceptance probability given by Eq.\ (\ref{p2}). We satisfy the
third requirement of equal wall-fluid interfacial tension in both
phases by exchanging wall area (and hence volume) between the two
boxes, while fixing both the wall separation distance $H$ in each
box as well as the total lateral area of both boxes
$A^\text{a}+A^\text{b}=A$ is constant, with $A^\text{a}$ and
$A^\text{b}$ the area of the subsystems a and b, respectively.  A
random walk in $\ln(A^\text{a}/A^\text{b})$ is then performed with
an acceptance probability given by
\begin{equation}
  \mathcal{P}=\text{min} \left\{1,  \mathcal{R}
  \exp[ -\beta ( U_\text{new}-U_\text{old})]  \right\},
  \label{p3.1}
\end{equation}
with
\begin{equation}
  \mathcal{R}= \left( \frac{A^\text{a}_\text{new}}{A^\text{a}_\text{old}}
  \right)^{N_\text{c}^\text{a}+N_\text{p}^\text{a}+1}
  \left( \frac{A^\text{b}_\text{new}}{A^\text{b}_\text{old}}
  \right)^{N_\text{c}^\text{b}+N_\text{p}^\text{b}+1}.
\end{equation}

We determine the fugacities of the colloids and of the polymers by
applying the GEMC version of the particle insertion method
\cite{BSDF89}
\begin{equation}
  z_k^\text{a,b} =  \left < \frac{V^\text{a,b}}{N^\text{a,b}_k+1}
  \exp[-\beta \Delta U ] \right >^{-1} \,  k=\text{c, p}
\end{equation}
where $a,b$ label the two simulation boxes and $\Delta U$ is the
energy defined by the acceptance rule of Eq.\ (\ref{p2}).  We
determine the densities of the two coexisting phases by sampling
histograms of the probability density $P(\eta_\text{c},\eta_\text{p})$
to observe the packing fractions $\eta_\text{p}$ and $\eta_\text{c}$
for the polymers and colloids, respectively.  The two maxima of
$P(\eta_\text{c},\eta_\text{p})$ correspond to the coexisting packing
fractions in the thermodynamic limit \cite{BSDF89}.  Statistical
uncertainties of the sampled quantities were determined by performing
three or four independent sets of simulations. The standard deviation
of the results from the simulation sets was used as the error
estimate.

\begin{figure}[htbp]
  \begin{center}
    \includegraphics[width=\mypic]{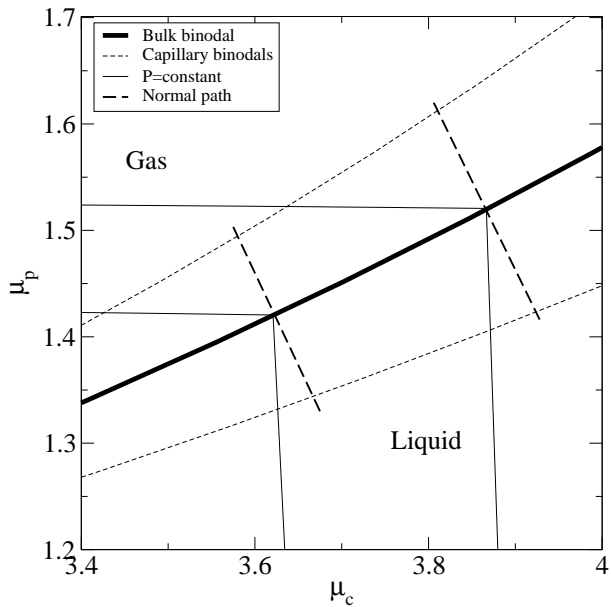}
\caption{Phase diagram of the AOV model with size ratio
      $q=\sigma_\text{p}/\sigma_\text{c}=1$ as a function of the
      colloidal chemical potential $\mu_\text{c}$ and the polymer
      chemical potential  $\mu_\text{p}$. The bulk binodal from free volume theory  (thick continuous curve)
      is shown along with isobaric lines (thin continuous curves) for two bulk reference points.  The thick dashed curves indicate the normal paths for the same reference points.
      The thin dashed curves are an illustration of the possible prediction of the Kelvin equation for the capillary binodals. For clarity we omitted the paths with constant polymer chemical potential. }
    \label{Fig4}
    \end{center}
\end{figure}

\section{Kelvin equation for binary mixtures}
\label{S:k1} In this Section, we derive expressions for the shift
in chemical potentials for the gas-liquid binodal of a binary
mixture confined between parallel plates assuming knowledge of
bulk quantities like the bulk coexisting densities, the chemical
potentials at bulk coexistence, and the liquid-wall and gas-wall
interfacial tensions, $\gamma_{\rm wl}$ and $\gamma_{\rm wg}$,
respectively.  Using a macroscopic picture we employ the grand
potential of the mixture confined in a slit of two parallel walls
with area $A$ and separation distance $h$  between the two walls
\footnote{We will discuss the relationship of $h$ to our model
parameter $H$ (see Sec.\ \ref{mod}) in more detail in Sec.\
\ref{S:k2}.} 
\begin{equation}
 \Omega(\mu_{\rm c},\mu_{\rm p}) = A h \omega(\mu_{\rm c},\mu_{\rm p}) +
   2 A \gamma(\mu_{\rm c},\mu_{\rm p},h),
 \label{EQomegaDivision}
\end{equation}
where $\mu_{\rm c}$ and $\mu_{\rm p}$ are the chemical potentials
of colloids and polymers, respectively, $\omega$ is the bulk grand
potential density (per unit volume), and $2\gamma$ is the
interfacial tension of the mixture and the plates at distance $h$.
For large $h$, we can approximate the latter quantity by twice the
interfacial tension of the mixture in contact with a single wall,
e.g., $2\gamma_{\rm w \alpha}(\mu_{\rm c},\mu_{\rm p})$, where $\alpha$=l,g are for the liquid and the gas
phase, respectively. The aim is to predict the grand potential in
the capillary given the knowledge of the thermodynamics of the
bulk at coexistence, i.e.\ at a statepoint specified through the
bulk values of the chemical potentials $\mu_{\rm c}^\coex$ and
$\mu_{\rm p}^\coex$. Hence, we can reexpress the chemical
potentials of both species for the confined fluid as $\mu_{\rm c}
= \mu_{\rm c}^\coex + \Delta \mu_{\rm c}$ and $\mu_{\rm p} =
\mu_{\rm p}^\coex + \Delta \mu_{\rm p}$. Quantities at coexistence
carry a superscript $\alpha=\rm l,g$, where $\rm l,g$ indicate
liquid and gas respectively.

The thermodynamic relations for the bulk densities of the colloids
and polymers, $\rho_c$ and $\rho_p$, respectively, read
\begin{equation}
 \rho_{\rm c} = -\left.\frac{\partial \omega}
  {\partial \mu_{\rm c}}\right|_{\mu_{\rm p}},
 \quad
 \rho_{\rm p} = -\left.\frac{\partial \omega}
  {\partial \mu_{\rm p}}\right|_{\mu_{\rm c}},
 \label{EQdensityFromGrandPotential}
\end{equation}
while the (excess) adsorption of the colloids and the polymers,
$\Gamma_c$ and $\Gamma_p$, are given by
\begin{equation}
 \Gamma_{\rm c} =
 \left. \frac{\partial \gamma_{\rm w \alpha}}{\partial \mu_{\rm c}}
 \right|_{\mu_{\rm p}}, \quad
 \Gamma_{\rm p} =
  \left. \frac{\partial \gamma_{\rm w \alpha}}{\partial
 \mu_{\rm p}}\right|_{\mu_{\rm c}}.
 \label{EQadsorptionFromTension}
\end{equation}
Using Eqs. (\ref{EQdensityFromGrandPotential}) and
(\ref{EQadsorptionFromTension}), we can perform a Taylor expansion
of the right hand side of Eq.\ (\ref{EQomegaDivision})
\begin{eqnarray}
 \frac{1}{Ah}\Omega(\mu_{\rm c},\mu_{\rm p}) \approx &&
 \omega(\mu_{\rm c}^\coex, \mu_{\rm p}^\coex)
 - \rho_{\rm c}^\alpha \Delta\mu_{\rm c} -
   \rho_{\rm p}^\alpha \Delta\mu_{\rm p}\nonumber\\ &&
 + \frac{2}{h} \Bigl[
 \gamma_{{\rm w}\alpha}
 + \Gamma_{\rm c}^\alpha \Delta \mu_{\rm c}
 + \Gamma_{\rm p}^\alpha \Delta \mu_{\rm p}
 \Bigr ],
\label{EQomegaExpansion}
\end{eqnarray}
where the bulk densities, $\rho_{\rm c}^{\alpha}$ and $\rho_{\rm p}^{\alpha}$,
and the (excess) adsorptions, $\Gamma_c^{\alpha}$ and
$\Gamma_{\rm p}^{\alpha}$, are evaluated in  one of the coexisting
phases
 $\alpha=\rm l,g$ at the statepoint given by
$\mu_{\rm c}^\coex$ and $\mu_{\rm p}^\coex$.

We next consider the capillary to be at phase coexistence, i.e.\ we
might envisage {\em two} capillaries in contact with each other, one
being filled with gas, the other being filled with liquid. Phase
equilibrium between both capillaries implies equality of the chemical
potential of each species and of the grand potential in both phases,
\begin{equation}
  \Omega_{\rm g}(\mu_{\rm c},\mu_{\rm p}) =
  \Omega_{\rm l}(\mu_{\rm c},\mu_{\rm p}),
 \label{EQomegaEquality}
\end{equation}
where $\Omega_{\rm g}$ and $\Omega_{\rm l}$ is the grand potential of
the gas and the liquid phase, respectively.
Using the approximation (\ref{EQomegaExpansion}) in
(\ref{EQomegaEquality}) yields
\begin{eqnarray}
&&\Bigl(\rho_{\rm c}^{\rm l} - \rho_{\rm c}^{\rm g} -
 \frac{2}{h}(\Gamma_{\rm c}^{\rm l} - \Gamma_{\rm c}^{\rm g})
\Bigr) \Delta \mu_{\rm c}  \nonumber\\
&+&
\Bigl(
 \rho_{\rm p}^{\rm l} - \rho_{\rm p}^{\rm g} -
 \frac{2}{h}(\Gamma_{\rm p}^{\rm l} - \Gamma_{\rm p}^{\rm g})
\Bigr) \Delta \mu_{\rm p}\quad\nonumber\\
&=&\frac{2}{h}\Bigl(
 \gamma_{\rm wl}(\mu_{\rm c}^\coex,\mu_{\rm p}^\coex) -
 \gamma_{\rm wg}(\mu_{\rm c}^\coex,\mu_{\rm p}^\coex)
\Bigr). \label{EQequationNine}
\end{eqnarray}

In the limit  $h$ is large compared with the
microscopic lengths and that the adsorptions remain finite, we can 
neglect  the terms proportional to $(\Gamma_i^{\rm
l}-\Gamma_i^{\rm g})/h$. Hence, Eq.\
(\ref{EQequationNine}) simplifies to 
\begin{eqnarray}
\bigl(
 \rho_{\rm c}^{\rm l} - \rho_{\rm c}^{\rm g}
\bigr) \Delta \mu_{\rm c}+
\bigl(
 \rho_{\rm p}^{\rm l} - \rho_{\rm p}^{\rm g}
\bigr) \Delta \mu_{\rm p}
=\frac{2}{h}\Bigl(
 \gamma_{\rm wl} -
 \gamma_{\rm wg}
\Bigr),
\label{EQequationTen}
\end{eqnarray}
which constitutes a single equation for the two unknown shifts in
the chemical potentials, $\Delta \mu_{\rm c}$ and $\Delta \mu_{\rm
p}$. In order to obtain a closed system of equations one requires
another assumption. We will present three different approaches in
the following. Each approach can be viewed as a different choice
of an ``optimal'' bulk reference state as illustrated in Fig.
\ref{Fig4}. A comparison with the numerical results will be
presented in Sec.\ \ref{S:k2}.

\subsection{Constant Polymer Chemical Potential}
The bulk reference state is chosen to be at
the same polymer chemical potential. This requirement is similar to
imposing the same temperature in a simple substance.  The optimal bulk
reference state with the same chemical potential of polymers leads to
\begin{equation}
 \Delta \mu_{\rm p} = 0.
 \label{EQmupConstant}
\end{equation}
Eqs.\ (\ref{EQequationTen}) and (\ref{EQmupConstant}) readily yield a
one-component Kelvin equation
\begin{equation}
 \Delta \mu_{\rm c} = \frac{2}{h}
 \frac{\gamma_{\rm wl}-\gamma_{\rm wg}}
      {\rho_{\rm c}^{\rm l}-\rho_{\rm c}^{\rm g}}.
\end{equation}
Clearly this result can be obtained with less labour by directly
dealing with the effective one-component system of colloids interacting with a
depletion potential, and rather serves as a illustration for the
validity of the reasoning leading to Eq.\ (\ref{EQequationTen}).

\subsection{Constant Pressure}
This reference state was used by Evans and Marini Bettolo Marconi in Ref. \cite{RE87}.
The bulk reference state is chosen to possess the same pressure.
In order to derive a corresponding condition consider the (finite
difference version of the) Gibbs-Duhem relation
\begin{equation}
 (S/V)\Delta T - \Delta P +
 \rho_{\rm c}\Delta \mu_{\rm c} + \rho_{\rm p}\Delta \mu_{\rm p} = 0,
 \label{EQgibbsDuhem}
\end{equation}
where $S$ is the entropy. Clearly, for athermal systems $\Delta
T=0$. We use (\ref{EQgibbsDuhem}) to compare the statepoint at
capillary coexistence with the reference statepoint at bulk
coexistence that possesses the same pressure, i.e.\ $\Delta P=0$,
hence
\begin{equation}
 \Delta \mu_{\rm p} = -
 \frac{\rho_{\rm c}^\alpha}{\rho_{\rm p}^\alpha} \Delta \mu_{\rm c},
 \quad \alpha=\rm g,l.
 \label{EQevansMBMreference}
\end{equation}
As shown in Fig. \ref{Fig4} the constant pressure paths (for which
the  Gibbs-Duhem relation is a good approximation close to the
bulk binodal)  have a discontinuity at bulk coexistence. To
predict the capillary phase behavior for statepoints that lie on
the gas side of the bulk gas-liquid binodal (hence to predict
capillary {\em condensation}), the correct reference state is the
coexisting {\em gas phase}, and we use
(\ref{EQevansMBMreference}) with $\alpha=\rm g$. Together with
(\ref{EQequationTen}) we obtain the classic result for capillary
condensation,
\begin{eqnarray}
  \Delta \mu_{\rm c} &=& \frac{2}{h}
  \frac{\gamma_{\rm wl}-\gamma_{\rm wg}}
       {\rho_{\rm c}^{\rm l} - (\rho_{\rm c}^{\rm g}/\rho_{\rm p}^{\rm g})
     \rho_{\rm p}^{\rm l}},
\label{EQkelvinEMBMcondensationC}\nonumber \\
  \Delta \mu_{\rm p} &=& \frac{2}{h}
  \frac{\gamma_{\rm wl}-\gamma_{\rm wg}}
       {\rho_{\rm p}^{\rm l}-
     (\rho_{\rm p}^{\rm g}/\rho_{\rm c}^{\rm g})\rho_{\rm c}^{\rm l}}.
\label{EQkelvinEMBMcondensationP}
\end{eqnarray}
Alternatively, predicting the capillary phase behavior for
statepoints that lie on the liquid side of the bulk gas-liquid
binodal (and  to predict capillary {\em evaporation}), we
choose the coexisting {\em liquid phase} as a reference. Eq.\
(\ref{EQevansMBMreference}) with $\alpha=\rm l$ together with Eq.\
(\ref{EQequationTen}) leads to the following result for capillary
evaporation
\begin{eqnarray}
  \Delta \mu_{\rm c} &=& -\frac{2}{h} \frac{\gamma_{\rm wl}-\gamma_{\rm wg}}
  {\rho_{\rm c}^{\rm g}-(\rho_{\rm c}^{\rm l}/\rho_{\rm p}^{\rm l})
    \rho_{\rm p}^{\rm g}},
\label{EQkelvinEMBMevaporationC} \nonumber \\
  \Delta \mu_{\rm p} &=& -\frac{2}{h} \frac{\gamma_{\rm wl}-\gamma_{\rm wg}}
  {\rho_{\rm p}^{\rm g}-(\rho_{\rm p}^{\rm l}/\rho_{\rm c}^{\rm l})
    \rho_{\rm c}^{\rm g}}.
\label{EQkelvinEMBMevaporationP}
\end{eqnarray}
The derivation presented here yields the same results as given in
Ref.\ \cite{RE87}, where a single capillary is considered in
contact with a bulk reservoir. We note that this procedure leads
to two different equations for the two phenomena of capillary condensation and evaporation.

\subsection{Normal Path Relation}
As a third and novel alternative we choose the reference state
such that the statepoint of interest lies on a path in the phase
diagram that crosses the bulk liquid-gas binodal in perpendicular
(or normal) direction in the plane of chemical potentials. This
implies optimizing proximity in the $\mu_c - \mu_p$ plane between
both statepoints.

The corresponding relation can readily be established starting from
the relation of the slope of the bulk binodal to the difference in
coexisting densities,
\begin{equation}
  \left.\frac{d \mu_{\rm c}}{d \mu_{\rm p}}\right|_{\rm coex} =
  -\frac{\rho_{\rm p}^{\rm l} - \rho_{\rm p}^{\rm g}}
  {\rho_{\rm c}^{\rm l} - \rho_{\rm c}^{\rm g}}.
\end{equation}
Hence a path normal to the bulk binodal is given by
\begin{equation}
  \left.\frac{d \mu_{\rm c}}{d \mu_{\rm p}}\right|_{\rm normal} =
  \frac{\rho_{\rm c}^{\rm l}-\rho_{\rm c}^{\rm g}}
  {\rho_{\rm p}^{\rm l} - \rho_{\rm p}^{\rm g}},
\end{equation}
from which we deduce the finite-difference version
\begin{equation}
 \Delta \mu_{\rm p} =
 \frac{\rho_{\rm p}^{\rm l} - \rho_{\rm p}^{\rm g}}
      {\rho_{\rm c}^{\rm l}-\rho_{\rm c}^{\rm g}}
      \Delta\mu_{\rm c},
\end{equation}
which, together with Eq.\ (\ref{EQequationTen}) yields
\begin{eqnarray}
  \Delta\mu_{\rm c} &=& \frac{2}{h} (\gamma_{\rm wl}-\gamma_{\rm wg})
   \frac{\rho_{\rm c}^{\rm l}-\rho_{\rm c}^{\rm g}}
   {(\rho_{\rm c}^{\rm l}-\rho_{\rm c}^{\rm g})^2  +
     (\rho_{\rm p}^{\rm l}-\rho_{\rm p}^{\rm g})^2}, \nonumber \\
  \Delta\mu_{\rm p} &=& \frac{2}{h} (\gamma_{\rm wl}-\gamma_{\rm wg})
   \frac{\rho_{\rm p}^{\rm l} - \rho_{\rm p}^{\rm g}}
   {(\rho_{\rm c}^{\rm l}-\rho_{\rm c}^{\rm g})^2 +
     (\rho_{\rm p}^{\rm l}-\rho_{\rm p}^{\rm l})^2},
     \label{EQkelvinnew}
\end{eqnarray}
valid both for capillary condensation and evaporation. As shown in
Fig. \ref{Fig4} the normal paths are symmetric with respect to the
gas and liquid side of the bulk binodal. The somewhat subtle
choice of the reference state, which is different for the two
phenomena in Sec. IV b, resulting in using either  Eq.
(\ref{EQkelvinEMBMcondensationP}) or Eq
(\ref{EQkelvinEMBMevaporationP}) is now avoided. The procedure
described here leads to one equation, which can be used for both
phenomena. This might be advantageous for mixtures where the
corresponding phases are less obvious, like e.g.\ confined liquid
crystals.

\begin{figure}[htbp]
  \begin{center}
    \includegraphics[width=\mypic]{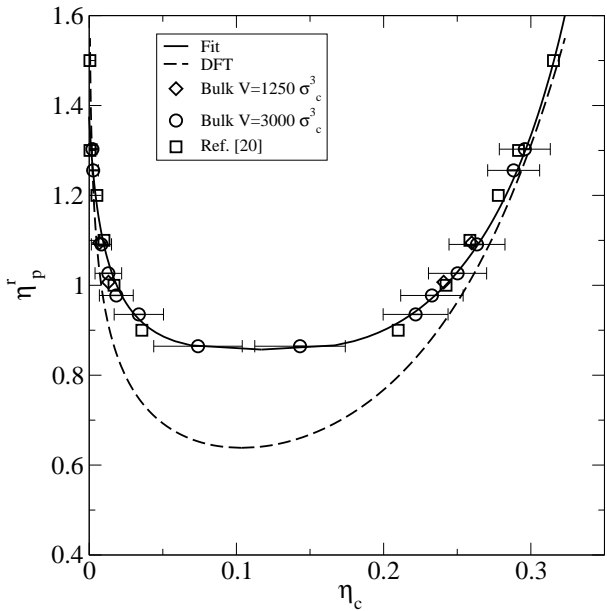}
    \caption{Bulk phase diagram of the AOV model with size ratio
      $q=\sigma_\text{p}/\sigma_\text{c}=1$ as a function of the
      colloidal packing fraction $\eta_\text{c}$ and the polymer
      reservoir packing fraction $\eta_\text{p}^\text{r}$.
      Coexistence is along the horizontal tie lines (not shown).
      Shown are GEMC simulation results for volumes
      $V=3000\sigma_\text{c}^3$ (circles) and
      $V=1250\sigma_\text{c}^3$ (diamonds), as well as the simulation
      results from Ref.\ \cite{MDRvR02} (squares).  We also display
      the result of the fit of Eqs.\ (\ref{scal}) and (\ref{diam}) (solid line) to the simulation data to and the binodal obtained from DFT, or, equivalently, free volume theory (dashed line).}
    \label{F:bulk}
  \end{center}
\end{figure}
\section{Results}
\label{res} In this section, we focus on colloid-polymer mixtures
with a size ratio $q=1$, a previously well-studied and also
experimentally accessible case. We have performed GEMC simulations
with $4 \times 10^8$ steps discarding the initial $10^8$ steps for
equilibration. The acceptance probability of particle displacement
was kept around 10\% to 20\%, the acceptance probability for
volume exchanges was around 50\%, while the acceptance probability
for transfer of particles was strongly dependent on the statepoint
and varied between 50\% and 5\% for the polymers and from 10\% to
less than 0.1\% for the colloids. The probability of a successful
particle transfer typically decreases with increasing particle
density. The statistical error of the fugacities serves as a good
indicator for the efficiency of the particle transfer between the
two boxes; an unreasonably high error indicates that the
simulation results are unreliable due to too low particle transfer
probabilities. Using this indicator the maximum packing fraction
that could be reached in our simulations was $\eta_\text{c}\approx
0.34$.

\begin{figure}[htbp]
  \begin{center}
    \includegraphics[width=\mypic]{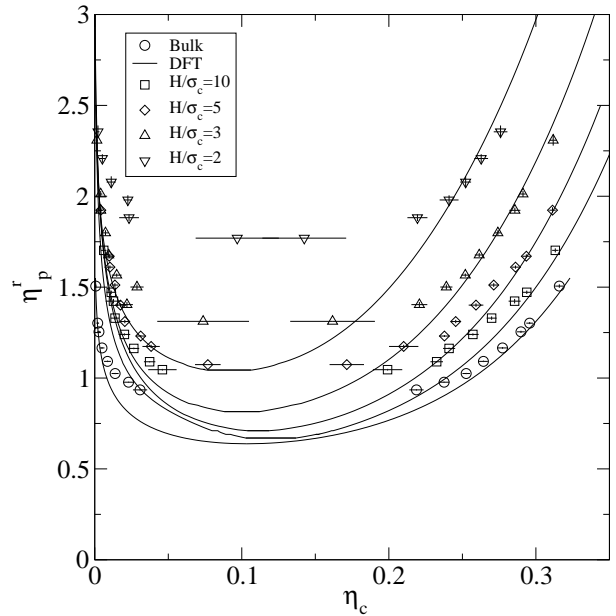}
  \end{center}
  \caption{Capillary phase diagram of the AOV model with size ratio
     $q=\sigma_\text{p}/\sigma_\text{c}=1$ confined between parallel
     hard walls with separation distance H/$\sigma_\text{c}$=2, 3, 5,
     10 and $\infty$ as a function of colloid packing fraction
     $\eta_\text{c}$ and polymer reservoir packing fraction
     $\eta_\text{p}^\text{r}$. Shown are results from simulation
     (symbols) and DFT (continuous lines).}
 \label{F:tot1}
\end{figure}

Fig.\ \ref{F:bulk} shows the bulk phase diagram obtained from
simulations with a simulation box of  volume $V=3000
\sigma_\text{c}^3$. Simulation runs with smaller system sizes
display negligible finite size effects.  In the
$\eta_\text{c}$-$\eta_\text{p}^\text{r}$ representation, the shape
of the binodal is similar to that of a simple fluid upon
identifying $\eta_\text{p}^\text{r}$ with inverse temperature. The
tielines, connecting the coexisting phases, are horizontal (not
shown) as $\eta_\text{p}^\text{r}$ possesses the same value in the
two  coexisting phases.  We have checked that our results agree
well with those obtained by Dijkstra and van Roij \cite{MDRvR02}
who performed simulations of  an effective one component system,
which was obtained by formally integrating out the degrees of
freedom of the polymers in the partition function.

To estimate the location of the critical point, we fitted the binodals
using the scaling law
\begin{equation}
  \eta_\text{c}^\text{l}-\eta_\text{c}^\text{g}=
  A \left( \frac{1}{(\eta_\text{p}^\text{r})_\text{crit}}-
  \frac{1}{\eta_\text{p}^\text{r}} \right)^\beta
\label{scal}
\end{equation}
and the law of rectilinear diameter
\begin{equation}
  \eta_\text{c}^\text{l}+\eta_\text{c}^\text{g}=
  2 (\eta_\text{c})_\text{crit} + B \left(   \frac{1}{(\eta_\text{p}^\text{r})_\text{crit}} - \frac{1}{\eta_\text{p}^\text{r}}  \right ),
\label{diam}
\end{equation}
where $\eta_\text{c}^\text{l}$ is the colloid packing fraction of the
coexisting liquid phase, $\eta_\text{c}^\text{g}$ is the colloid
packing fraction of the coexisting gas phase and the subscript $crit$
indicates the value at the critical point. $A$ and $B$ are two free
parameters determined from the fit, and $\beta=0.32$ is the
three-dimensional Ising critical exponent. We used the standard
functional form of the two laws \cite{DFBS96} but replaced the
temperature by the inverse of the polymer reservoir packing
fraction. The continuous curve in Fig.\ \ref{F:bulk} is the result of
the fit of Eqs.\ (\ref{scal}) and (\ref{diam}). The fit is remarkably
good, but we point out that this only gives an estimate of the
critical point.  To get a more precise value of the critical packing
fractions, it would be necessary to carry out simulations in a region
of the phase diagram much closer to the critical point than is
possible with GEMC.

In Fig.\ \ref{F:bulk}, we also compare our results with those obtained
from density functional theory (DFT). We use the approximation for the
Helmholtz excess free energy for the AOV model as given in
\cite{Schmidt2000}. For given external potential, the density
functional is numerically minimized using a standard iteration
procedure.  The discrepancies between theory and simulation can be
understood by considering that the DFT for homogeneous (bulk) fluid
states of the AOV model is equivalent to the free volume theory of
Lekkerkerker \et \cite{HNWL+92}.  Dijkstra \et \cite{MDJMB99} showed
that this theory is equivalent to a first order Taylor expansion of
the free energy around $\eta_\text{p}^\text{r}=0$
\begin{eqnarray}
& &\beta F(N_\text{c},V,\eta_\text{p}^\text{r})  =   \beta F(N_\text{c},V,\eta_\text{p}^\text{r}=0) + \nonumber \\
&+&\int_\text{0}^{\eta_\text{p}^\text{r}} d (\eta_\text{p}^\text{r})' \left (\frac{\partial \beta F
(N_\text{c},V,(\eta_\text{p}^\text{r})')}{\partial (\eta_\text{p}^\text{r})'} \right )_{(\eta_\text{p}^\text{r})'}
\nonumber \\
&\simeq& \beta F(N_\text{c},V,\eta_\text{p}^\text{r}=0)- \frac{6}{\pi \sigma_\text{p}^3}
\eta_\text{p}^\text{r} < V_\text{free}
>_{\eta_\text{p}^\text{r}=0} \, , \label{fvt}
\end{eqnarray}
neglecting terms ${\cal O}((\eta_\text{p}^\text{r})^2)$ and where
$\langle V_\text{free} \rangle_{\eta_\text{p}^\text{r}=0}$ is the
free volume available for the polymer in the pure hard sphere
reference system. It is evident in Fig.\ \ref{F:bulk} that the
theory (dashed curve) performs better at high $\eta_\text{c}$
where the system is so crowded that it resembles the reference
hard sphere system and $\langle V_\text{free}
\rangle_{(\eta_\text{p}^\text{r})'=0} \simeq \langle V_\text{free}
\rangle_{(\eta_\text{p}^\text{r})'=\eta_\text{p}^\text{r}}$.  For
very small $\eta_\text{c}$ the free volume is close to the total
volume of the system $V$ for both the pure hard sphere reference
system and the actual mixture. For high $\eta_\text{p}^\text{r}$,
the gas-liquid coexistence is very broad and quantitatively
well-predicted by the theory. The critical point of the AOV
mixture for size ratio $q=1$ is in the region of $\eta_\text{c}
\sim 0.1$ and the theory underestimates the critical value of
$\eta_\text{p}^\text{r}$. Furthermore, the discrepancy in location
of the critical point arises from the mean field critical exponent
of the theory against the 3D Ising critical exponent of the
simulation \cite{RLCV04}.

\begin{figure}[htbp]
  \begin{center}
    \includegraphics[width=\mypic]{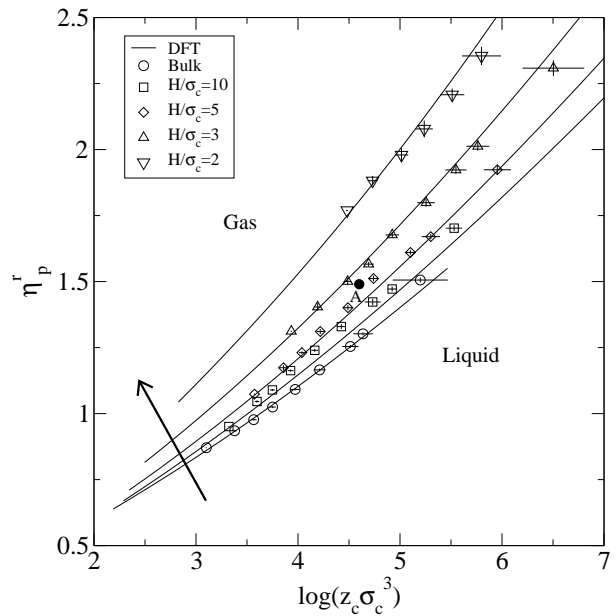}
    \caption{The same as Fig.\ \ref{F:tot1}, but as a function of
      colloid fugacity $z_\text{c} \sigma_\text{c}^3$ and polymer
      reservoir packing fraction $\eta_\text{p}^\text{r}$. Statepoint
      A is in the gas region of the phase diagram for wall separation
      distances $H/\sigma_\text{c}=\infty$, 10, and 5 and in the
      liquid region of the phase diagram for wall separation distances
      $H/\sigma_\text{c}=3$ and 2. The arrow indicates the direction
      of the binodal shift upon increasing confinement (decreasing
      values of $H/\sigma_\text{c}$) of the mixture between parallel
      hard walls.}
    \label{F:tot2}
  \end{center}
\end{figure}

\begin{table}[tb]
\begin{center}
\begin{tabular}{|c|c|c|c|c|}
\hline
$H/\sigma_\text{c}$ & $(\eta_\text{p}^\text{r})_\text{crit}$ & $(\eta_\text{c})_\text{crit}$ &  DFT $(\eta_\text{p}^\text{r})_\text{crit}$ & DFT $(\eta_\text{c})_\text{crit}$ \\
\hline
   $\infty$  & 0.86(1) &  0.117(2) &   0.638 & 0.103 \\
   10        & 1.00(1) &  0.124(1) &   0.670 & 0.120 \\
   5         & 1.09(1) &  0.123(2) &   0.710 & 0.111 \\
   3         & 1.27(2) &  0.116(3) &   0.815 & 0.100 \\
   2         & 1.76(2) &  0.119(1) &   1.044 & 0.091 \\
\hline
\end{tabular}
\end{center}
\caption{ Capillary critical points of the AOV model between parallel
  hard walls with separation distance $H/\sigma_\text{c}=\infty, 10,
  5, 3,$ and 2 obtained from the fit of Eqs.\ (\ref{scal}) and
  (\ref{diam}), and from DFT.  } \label{tabcri}
\end{table}

\subsection{Phase diagrams}
\label{phds}
First we present the results for the colloid-polymer mixtures confined
between two smooth, planar hard walls at distance $H$.  In Fig.\
\ref{F:tot1} we show a set of phase diagrams for
$H/\sigma_\text{c}=\infty$ (bulk), 10, 5, 3, and 2, in the
$\eta_\text{c}$-$\eta_\text{p}^\text{r}$ representation. Upon
decreasing the plate separation distance $H/\sigma_\text{c}$, the
critical value of $\eta_\text{p}^\text{r}$ shifts to higher values, in
accordance with the decrease in critical temperature of simple
fluids. The theoretical binodals agree well with those from
simulation, except close to the critical point.  The theory
underestimates for all plate separations the critical value of
$\eta_\text{p}^\text{r}$, as it does for the bulk system
\cite{MDRvR02}.  We observe that the deviation increases upon
decreasing $H/\sigma_\text{c}$.  In Tab.\ \ref{tabcri} we show the
critical packing fractions obtained from the fit of Eqs.\ (\ref{scal})
and (\ref{diam}) and from DFT.  We used the 3-dimensional Ising
critical exponent for all wall separations. Although recent studies
\cite{vink2005} suggests a critical behavior for small wall
separations that is neither three-dimensional, nor two-dimensional,
the difference is likely to be negligible at the level of precision of
our GEMC simulations.


In Fig.\ \ref{F:tot2} we show a set of phase diagrams for
$H/\sigma_\text{c}=\infty$, 10, 5, 3, and 2 in the
$z_\text{c}$-$\eta_\text{p}^\text{r}$ representation.  The
coexistence gap in colloid packing fractions collapses to a line
since two phases at coexistence possess the same colloid fugacity.
Note that the system is in the gas phase for fugacity $z_\text{c}<
z_\text{c}^\text{sat}$, while it is in the liquid phase for
$z_\text{c}> z_\text{c}^\text{sat}$, where $z_\text{c}^\text{sat}$
denotes the colloid fugacity at bulk coexistence. Statepoint A is
gas-like for $H/\sigma_\text{c} \geq 5$, but is liquid-like for
$H/\sigma_\text{c} \leq 3$. Hence, planar slits with
$H/\sigma_\text{c} \leq 3$ are filled with liquid phase, while the
bulk reservoir is in the gas phase, proving the occurrence of
capillary condensation upon reducing  the width of the slit that
is in contact with a bulk gas. 

\begin{figure}[htbp]
  \begin{center}
    \includegraphics[width=\mypic]{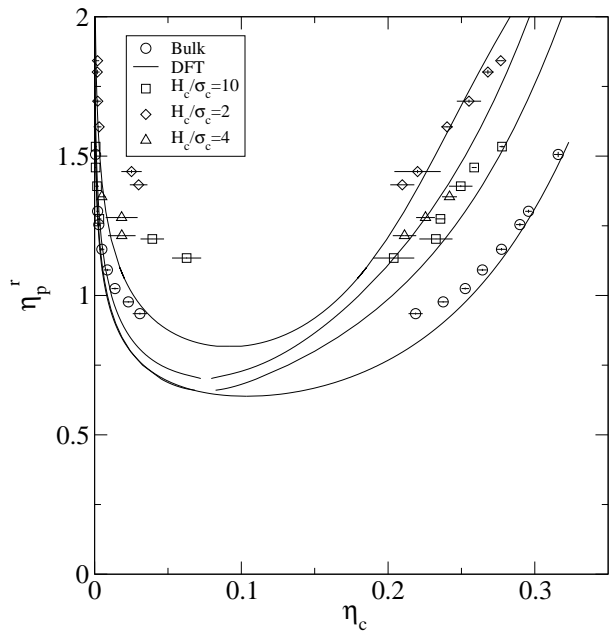}
    \caption{Capillary phase diagram showing the gas-liquid binodal for the AOV
      model with $q=\sigma_\text{p}/\sigma_\text{c}=1$ between two
      parallel semi-permeable walls with separation distances
      $H_\text{c}/\sigma_\text{c}$=2, 4, 10, $\infty$ as a function of
      colloid packing fraction $\eta_\text{c}$ and polymer reservoir
      packing fraction $\eta_\text{p}^\text{r}$. Shown are results
      from simulations (symbols) and DFT (lines).}
    \label{F:stot1}
  \end{center}
\end{figure}

\begin{figure}[htbp]
  \begin{center}
    \includegraphics[width=\mypic]{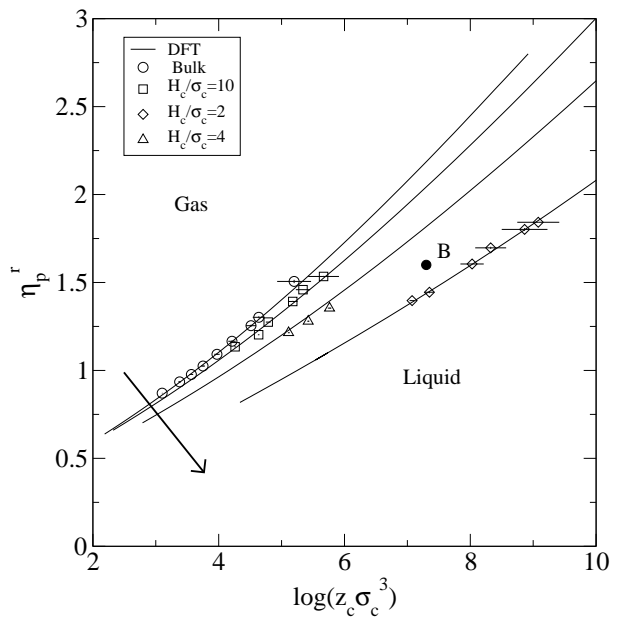}
    \caption{The same as Fig.\ \ref{F:stot1}, but as a function of
      colloid fugacity $z_\text{c} \sigma_\text{c}^3$ and polymer
      reservoir packing fraction $\eta_\text{p}^\text{r}$. The arrow
      indicates the direction of the shift of the binodal upon
      increasing confinement of the mixture (decreasing values of
      $H_\text{c}/\sigma_\text{c}$) between two parallel semi-permeable walls.}
    \label{F:stot2}
  \end{center}
\end{figure}

We now turn our attention to colloid-polymer mixtures confined between
two smooth, planar semi-permeable walls at distance $H_\text{c}$.  In
Fig.\ \ref{F:stot1} we show a set of phase diagrams for
$H_\text{c}/\sigma_\text{c}=\infty$, 10, 4, and 2 in the
$\eta_\text{c}$-$\eta_\text{p}^\text{r}$ representation. Upon
increasing the confinement (via reduction of
$H_\text{c}/\sigma_\text{c}$), the critical value of
$\eta_\text{p}^\text{r}$ shifts to higher values. The trend is similar
to the behavior of the slit with hard walls, although we find a
smaller shift of the critical point (see Tab. \ref{stabcri}).  In
Fig.\ \ref{F:stot2} we show a set of phase diagrams for
$H/\sigma_\text{c}=\infty, 10, 4,$ and 2 in the
$z_\text{c}$-$\eta_\text{p}^\text{r}$ representation.  Statepoint B is
liquid-like for $H_\text{c}/\sigma_\text{c} \geq 10$, but is gas-like
for $H_\text{c}/\sigma_\text{c} \leq 2$. Hence, planar slits with
$H_\text{c}/\sigma_\text{c} \leq 2$ are filled with gas, while the
bulk reservoir is in the liquid phase, indicating the occurrence of
capillary evaporation.

\begin{table}[tb]
\begin{center}
\begin{tabular}{|c|c|c|c|c|}
\hline
$H/\sigma_\text{c}$ & $(\eta_\text{p}^\text{r})_\text{crit}$ & $(\eta_\text{c})_\text{crit}$ &  DFT $(\eta_\text{p}^\text{r})_\text{crit}$ & DFT $(\eta_\text{c})_\text{crit}$ \\
\hline
 $\infty$  & 0.86(1)   &  0.117(2)  &  0.638 & 0.103 \\
   10      & 1.09(2)   &  0.13(1)   &  0.660 & 0.075  \\
   4       &  1.11(4)  &  0.10(3)   &  0.699 & 0.076 \\
   2       &   1.29(1) &  0.11(4)   &  0.818 & 0.092 \\
\hline
\end{tabular}
\end{center}
\caption{Capillary critical points of the AOV model between parallel
  semi-permeable plates with separation distance
  $H/\sigma_\text{c}=\infty, 10, 4,$ and 2 obtained from the fit of
  Eqs.\ (\ref{scal}) and (\ref{diam}), and from DFT. }
\label{stabcri}
\end{table}

\subsection{Structure at coexistence}
\label{str} We next analyze the density profiles of both species
at capillary coexistence of gas and liquid phases. Such fluid
states are translationally invariant against lateral
displacements, and the density distributions (of both species)
depend solely on the (perpendicular) distance from the walls. We
compare theoretical and simulation results for coexistence states
at the same values for $\eta_\text{p}^\text{r}$. In practice, we
have used the result for $\eta_\text{p}^\text{r}$ from the
simulations, and have calculated the corresponding DFT profiles.
The value for $z_\text{c}$ used in the DFT calculations was
adjusted according to the respective theoretical capillary
binodal. Recall that the quantitative differences in results for
the capillary binodals from simulation and theory are small.

\begin{figure}[htbp]
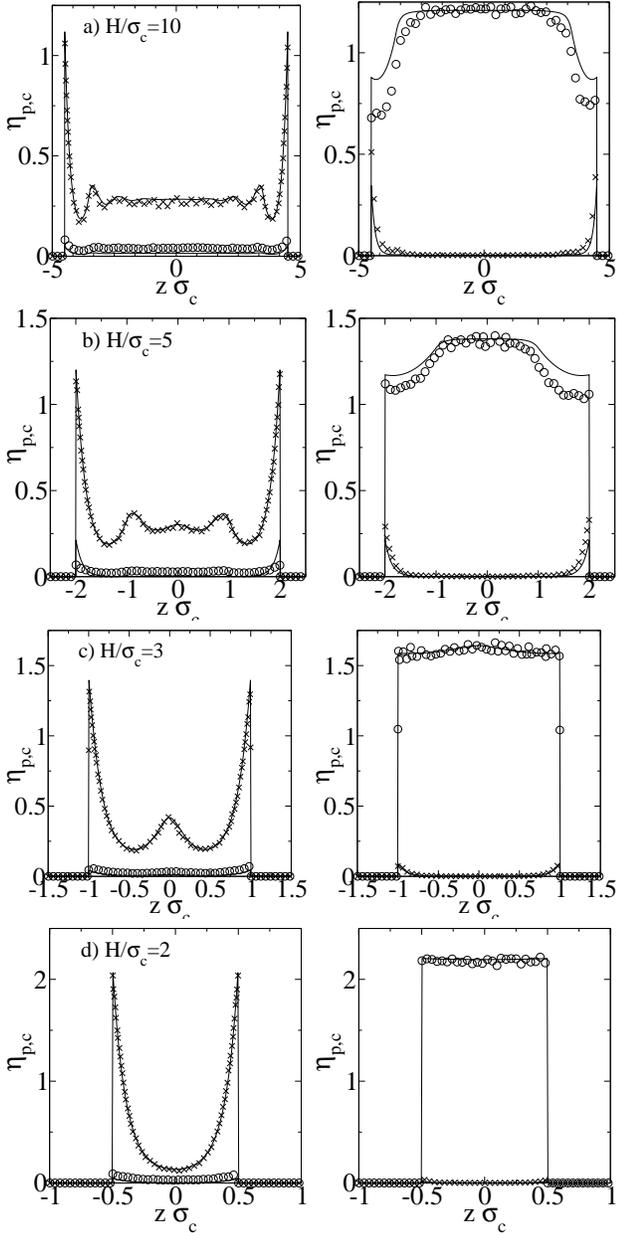

 \includegraphics[width=4.0cm] {fig10a1.eps}
 \includegraphics[width=4.0cm] {fig10a2.eps}
 \includegraphics[width=4.0cm] {fig10b1.eps}
 \includegraphics[width=4.0cm] {fig10b2.eps}
 \includegraphics[width=4.0cm] {fig10c1.eps}
 \includegraphics[width=4.0cm] {fig10c2.eps}
 \includegraphics[width=4.0cm] {fig10d1.eps}
 \includegraphics[width=4.0cm] {fig10d2.eps}
    \caption{Density profiles of the coexisting liquid (left) and gas
      (right) phases of the AOV model  confined between parallel hard walls with
      separation distance: a) H/$\sigma_\text{c}$=10, at
      $\eta_\text{p}^\text{r}$=1.23(1),$z_\text{c}
      \sigma_\text{c}^3$=66.3(1), b) H/$\sigma_\text{c}$=5 at
      $\eta_\text{p}^\text{r}$=1.39(1), $z_\text{c}
      \sigma_\text{c}^3$=99.4(3), c) H/$\sigma_\text{c}$=3, at
      $\eta_\text{p}^\text{r}$=1.68(1), $z_\text{c}
      \sigma_\text{c}^3$=128(2), and d) H/$\sigma_\text{c}$=2, at
      $\eta_\text{p}^\text{r}$=2.23(1), $z_\text{c}
      \sigma_\text{c}^3$=228(10). Shown are results from simulations
      for the density profiles of the colloids (crosses) and polymers
      (circles), along with results from DFT (full lines). }
 \label{F:prof2}
\end{figure}

The results for a slit of hard walls are shown in Fig.\
\ref{F:prof2}. The colloidal profiles in the liquid phase display
strong layering at either wall. For wall separation
$H/\sigma_\text{c}=10$, at $\eta_\text{p}^\text{r}=1.23 \pm 0.01$,
and $z_\text{c} \sigma_\text{c}^3=66.3 \pm 0.1$ these oscillations
decay to flat, bulk-like behavior in the center of the slit. For
smaller wall separations, namely $H/\sigma_\text{c}=5$, at
$\eta_\text{p}^\text{r}=1.39 \pm 0.01$, and $z_\text{c}
\sigma_\text{c}^3=99.4 \pm 0.3$, $H/\sigma_\text{c}=3$, at
$\eta_\text{p}^\text{r}=1.68 \pm 0.01$, and $z_\text{c}
\sigma_\text{c}^3=128 \pm 2$, and $H/\sigma_\text{c}=2$, at
$\eta_\text{p}^\text{r}=2.23 \pm 0.01$, and $z_\text{c}
\sigma_\text{c}^3=228 \pm 10$, we observe the presence of 5, 3, and
2 well-defined layers of particles, respectively.  Although their
density is much lower, the polymers in the liquid phase display
similar behavior.  The layering is weaker, but we can observe that
a maximum in the colloid profile corresponds also to a maximum in
the polymer profile. This result suggests that for such low
concentrations polymers behave as hard spheres as packing effects
are concerned.  In the gas phase, for wall separations
$H/\sigma_\text{c}=10$ and 5, we find strong adsorption of the
colloids at both walls,  
and a tendency of the polymers to desorb from the
walls. In the center of the slit almost no colloids are present
and the polymers display flat density profiles with a packing
fraction very similar to the polymer reservoir packing fraction.
For wall separation distances $H/\sigma_\text{c}=3$ and 2, we
observe an almost flat polymer density profile, while the density
of the colloids is very low throughout the slit. Different from
the liquid profiles, a maximum in the colloidal profile
corresponds  to a minimum of the polymer profiles.

\begin{figure}[htbp]
\includegraphics[width=4.0cm] {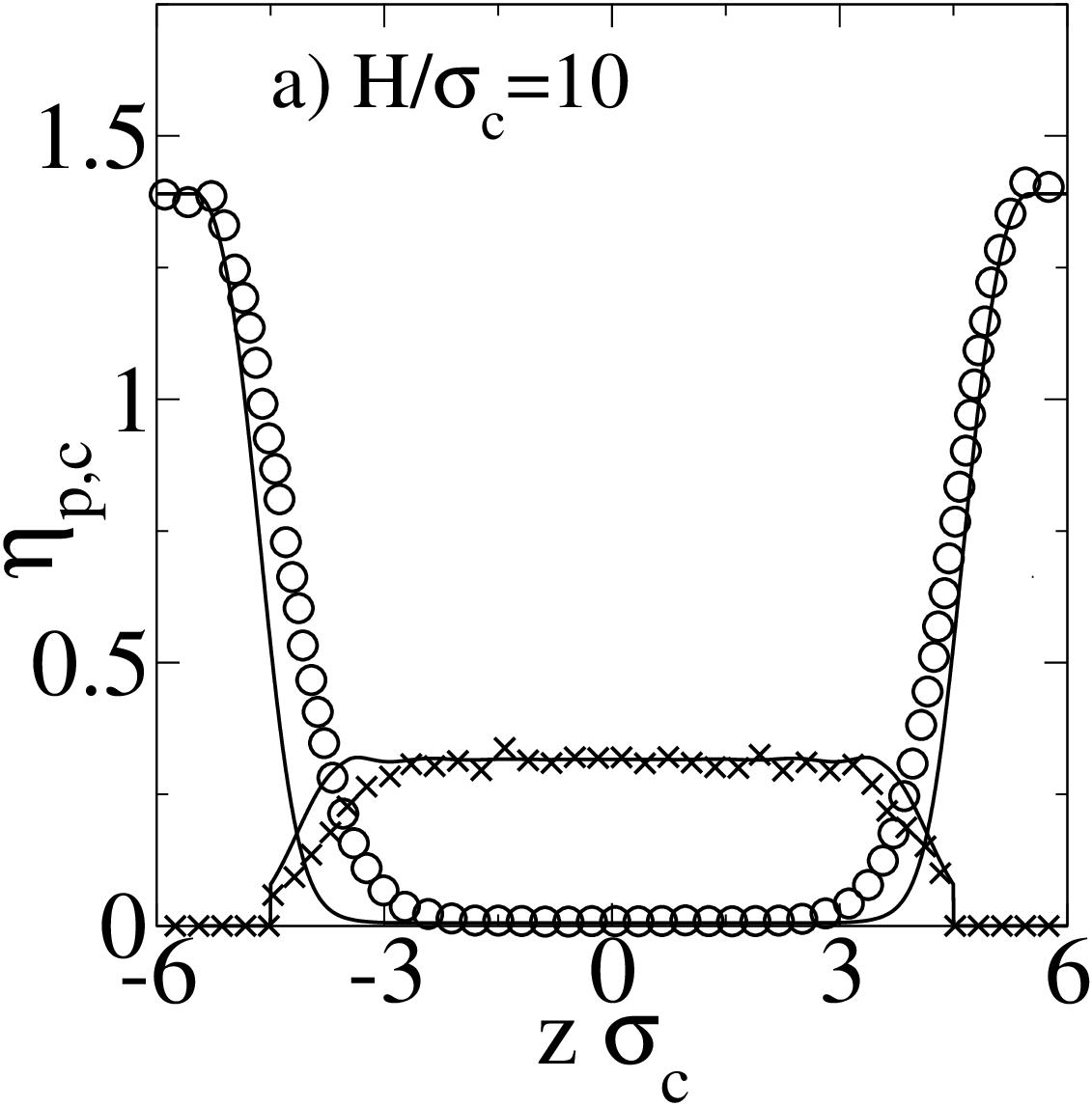}
\includegraphics[width=4.0cm] {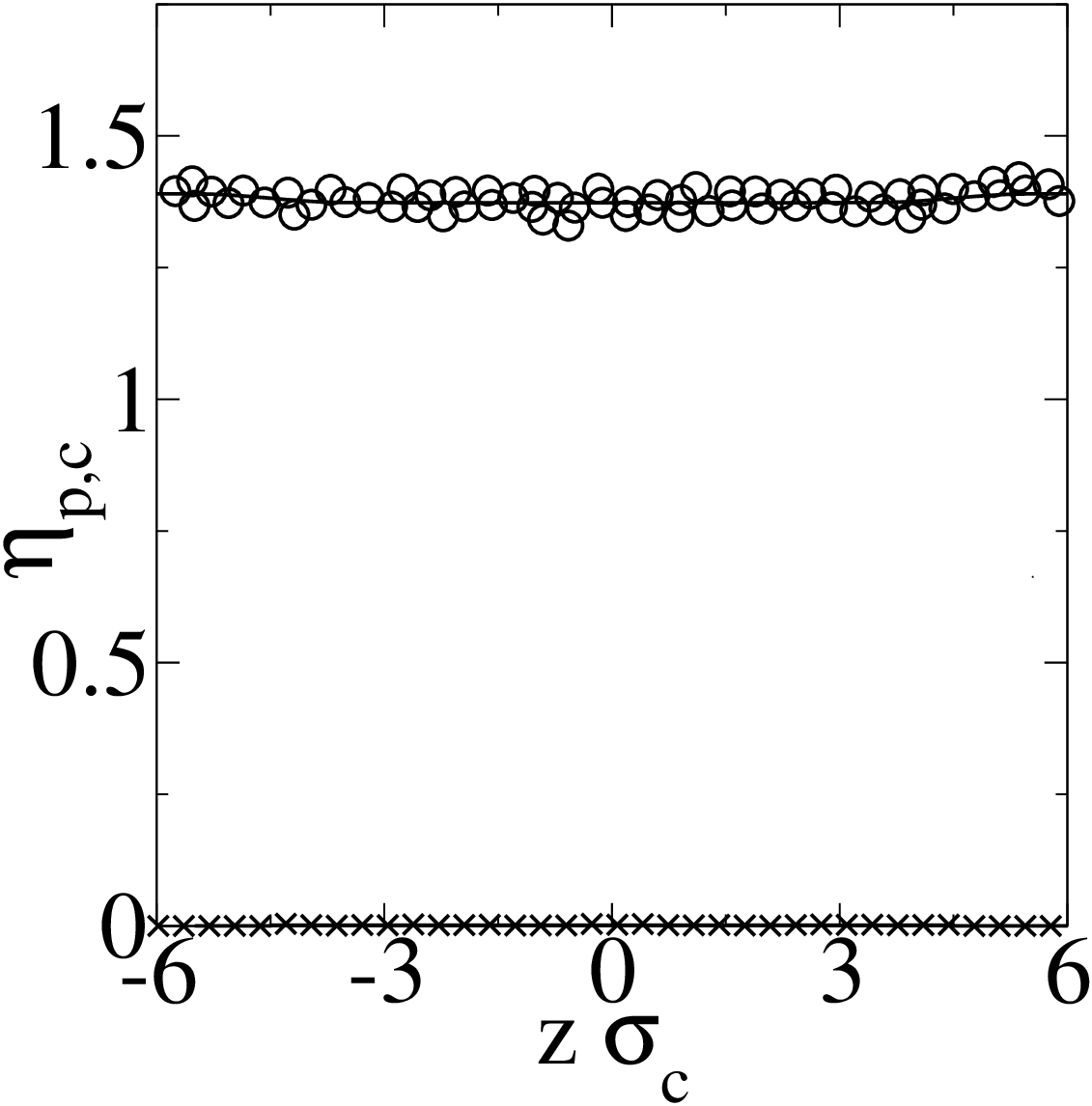}
 \includegraphics[width=4.0cm] {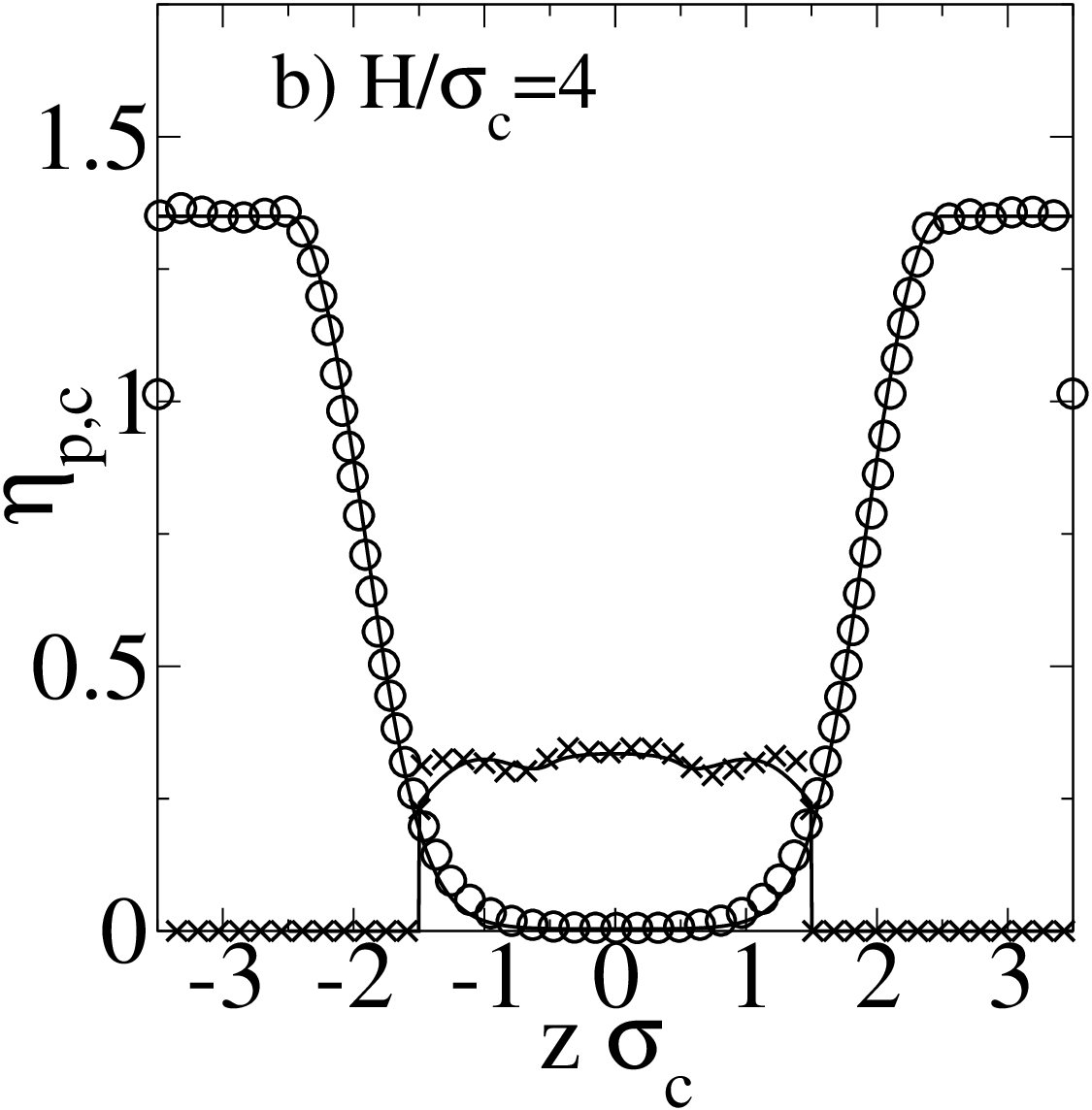}
\includegraphics[width=4.0cm] {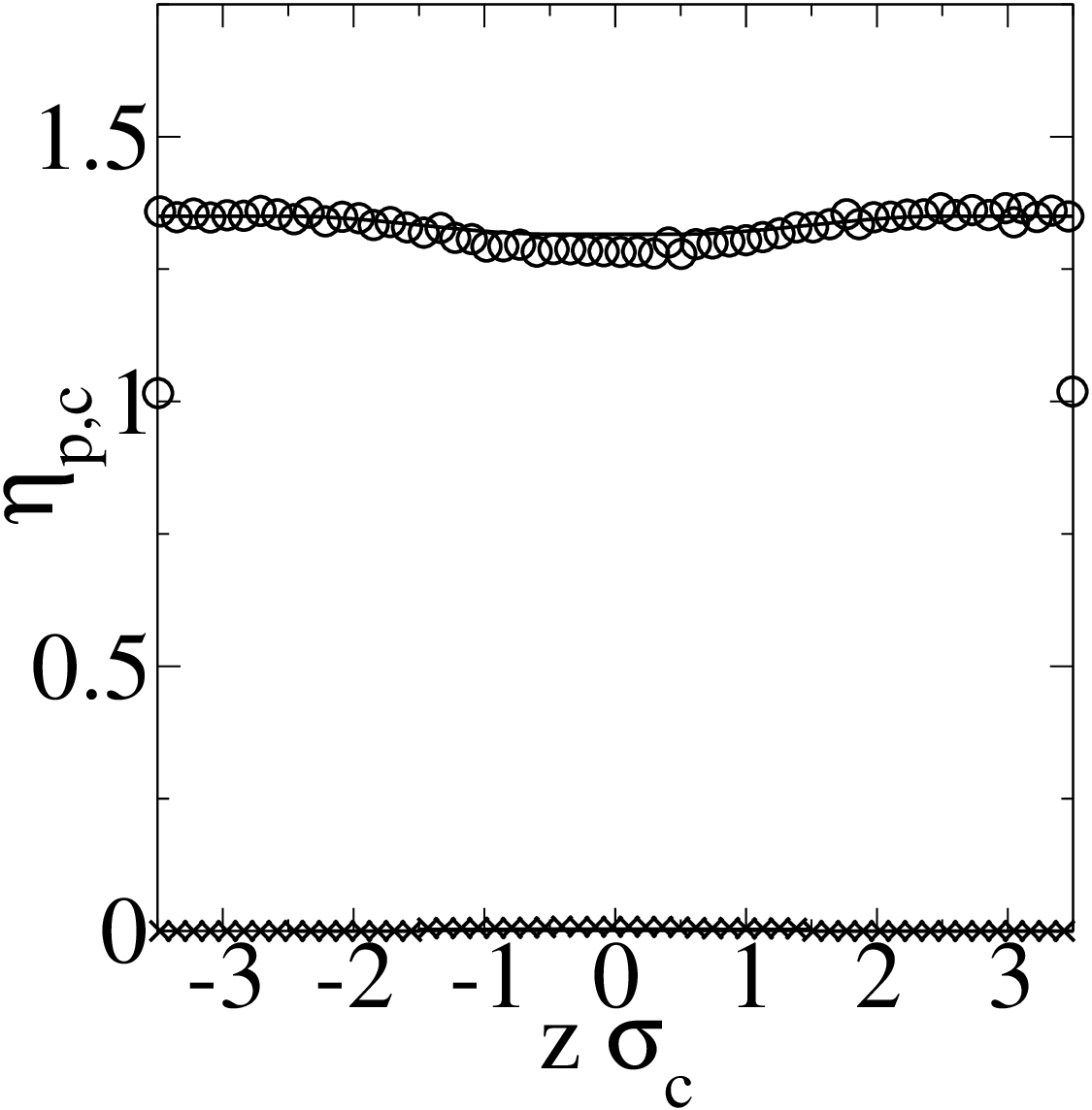}
 \includegraphics[width=4.0cm] {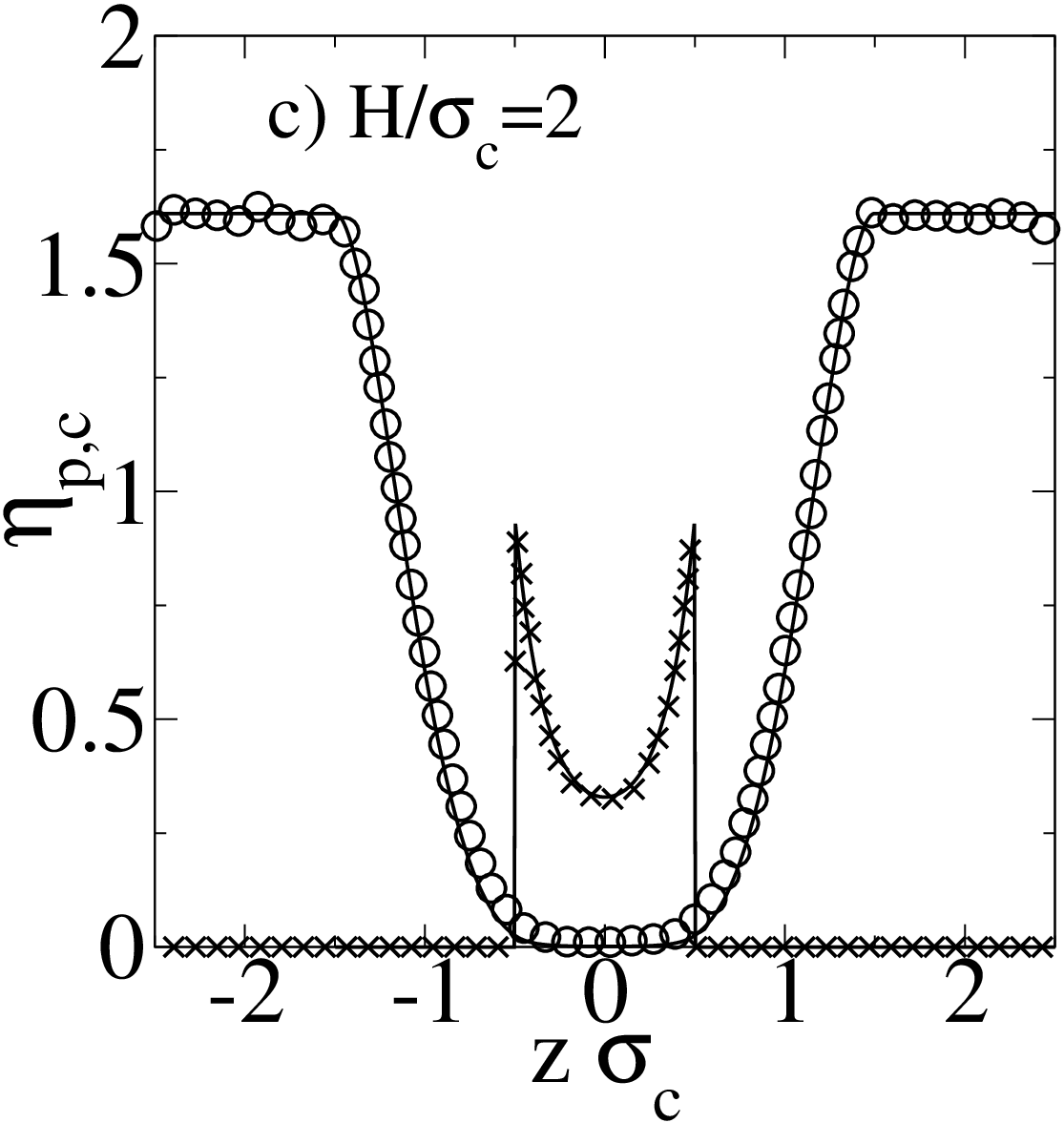}
\includegraphics[width=4.0cm] {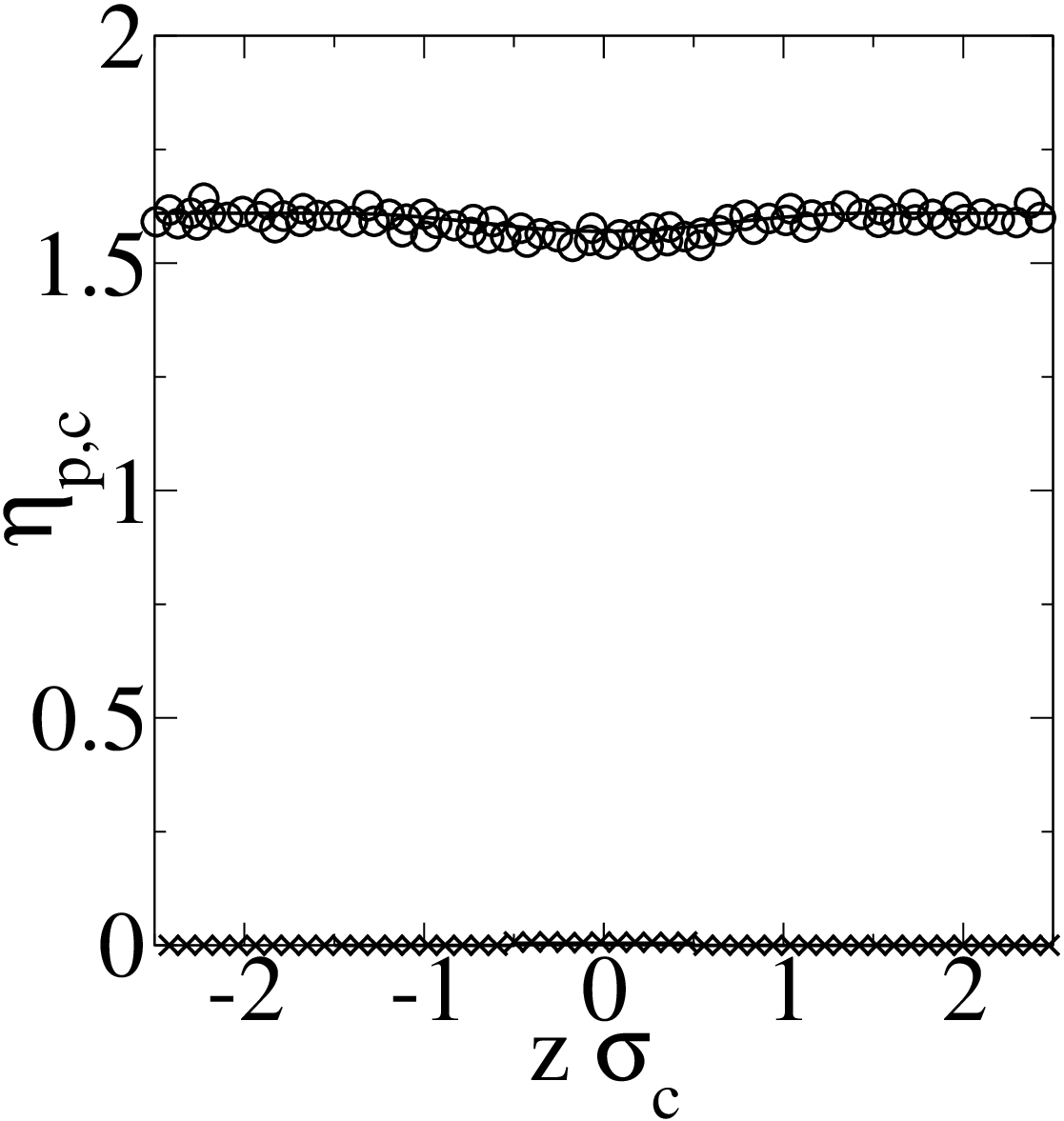}
 \caption{Density profiles of the coexisting liquid (left) and gas
    (right) phases of the AOV model confined between two parallel
    semi-permeable walls with varying separation distance: a)
    H$_\text{c}$/$\sigma_\text{c}$=10, at
    $\eta_\text{p}^\text{r}$=1.39(1), $z_\text{c}
    \sigma_\text{c}^3$=167(5), b) H$_\text{c}$/$\sigma_\text{c}$=4, at
    $\eta_\text{p}^\text{r}$=1.35(1), $z_\text{c}
    \sigma_\text{c}^3$=318(10), and c)
    H$_\text{c}$/$\sigma_\text{c}$=2, at
    $\eta_\text{p}^\text{r}$=1.61(1), $z_\text{c}
    \sigma_\text{c}^3$=2950(200).  Shown are results from simulations
    for the density profiles of the colloids (crosses) and polymers
    (circles), and results from DFT (full lines).}
 \label{F:sprof2}
\end{figure}

Fig.\ \ref{F:sprof2} displays density profiles for the slit of
semi-permeable walls.  For wall separation distance of
$H_\text{c}/\sigma_\text{c}=10$, at $\eta_\text{p}^\text{r}=1.39\pm
0.01$, and $z_\text{c} \sigma_\text{c}^3=167 \pm 5$, we clearly
observe for the liquid state points the presence of a gas layer between the wall and the liquid
phase centered in the slit.  For wall separation
$H_\text{c}/\sigma_\text{c}=4$, at $\eta_\text{p}^\text{r}=1.35 \pm
0.01$, and $z_\text{c} \sigma_\text{c}^3=318 \pm 10$ the gas layers at
the walls disappear and indications of layering effects appear.  For
wall separation $H_\text{c}/\sigma_\text{c}=2$, at
$\eta_\text{p}^\text{r}=1.61 \pm 0.01$, and $z_\text{c}
\sigma_\text{c}^3=2950 \pm 200$ the colloid density profile displays
very significant peaks at both walls. Moreover, we do find layering at
larger wall separations, for statepoints well inside the liquid phase.
We will discuss this in more detail in the next section. In the gas phase, the density of 
colloids is very low throughout the slit, while the polymer density profile is almost flat with a packing fraction close to the polymer reservoir packing fraction. 

The comparison between DFT and simulation indicates good agreement of
results from both approaches.  Differences in structure can be traced
back to differences in the phase diagrams. For fixed
$\eta_\text{p}^\text{r}$ the DFT predicts higher colloid densities and
smaller polymer densities, and these differences are reflected by the
discrepancies in the profiles. For wall separations where the
statepoint is close to the critical point the agreement is worse,
especially close to the walls.  Such discrepancies between simulation
and DFT were previously reported in studying the wall-fluid tension
and the adsorption of colloid-polymer mixtures \cite{Fortini2005a}.

\subsection{Structure off-coexistence}
\begin{figure}[htbp]
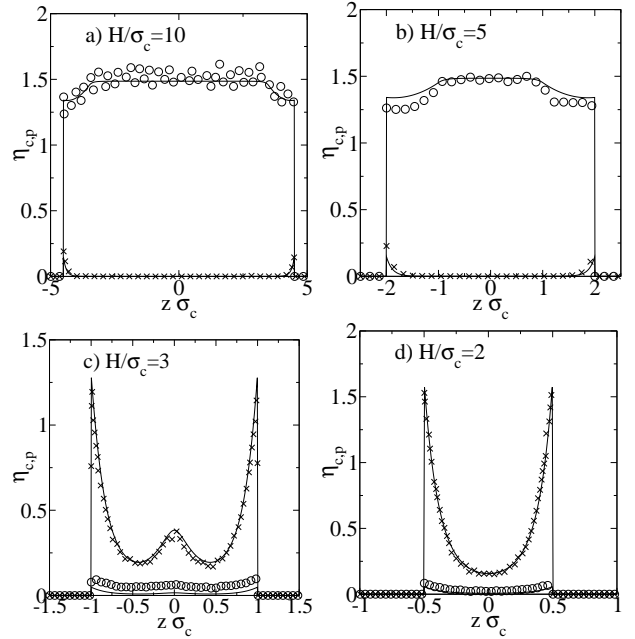

\begin{center}
\includegraphics[width=4.0cm] {fig12a.eps}
\includegraphics[width=4.0cm] {fig12b.eps}
\includegraphics[width=4.0cm] {fig12c.eps}
\includegraphics[width=4.0cm] {fig12d.eps}
\end{center}
\caption{Density profiles of the AOV model with $q=1$ between parallel
  hard walls with varying separation distance: a)
  $H/\sigma_\text{c}$=10, b) $H_\text{c}/\sigma_\text{c}$= 5, c)
  $H_\text{c}/\sigma_\text{c}$=3, and d)
  $H_\text{c}/\sigma_\text{c}$=2 at statepoint A of the phase diagram
  of Fig.\ \ref{F:tot2}, i.e., for polymer reservoir packing fraction
  $\eta_\text{p}^\text{r}=1.49$ and colloid fugacity $\ln(z_\text{c}
  \sigma_\text{c}^3)=4.6$. Shown are results from simulations for the
  density profiles of the colloids (crosses) and polymers (circles),
  and results from DFT (full lines).}
\label{F:prof1}
\end{figure}

We next consider density profiles for a fixed statepoint
off-coexistence. For slits with hard walls we chose statepoint A
($\eta_\text{p}^\text{r}$=1.49 and $\ln(z_\text{c}
\sigma_\text{c}^3)$=4.6) of Fig.\ \ref{F:tot2}, that lies in the
stable gas region of the bulk phase diagram.  We carried out
simulations for wall separation distances $H/\sigma_\text{c}=10, 5,
3,$ and 2.  Fig.\ \ref{F:prof1} shows that for wall separations
$H/\sigma_\text{c}=10$ and 5 the slit is filled with gas.  However,
for wall separation of $H/\sigma_\text{c}=3$ we observe that the
capillary fills with liquid.  Hence, for this particular statepoint,
the critical wall separation distance for capillary condensation lies
between 3 and 5 colloid diameters, consistently with the findings of
Sec.\ \ref{phds}.  Reducing the wall separation to
$H/\sigma_\text{c}=2$, the liquid phase remains stable.  The density
profiles in the gas phase possess adsorption peaks in the colloid
profile, and corresponding desorption peaks in the polymer
profiles. In the liquid phase we observe strong layering of the
colloids, and, to a lesser extent, of the polymers.  The agreement
between simulation and DFT results is good. The differences seem to be
related to the vicinity of statepoint A to the critical point.  The
critical point for wall separation $H/\sigma_\text{c}=3$, and 5, is
much closer to statepoint A than the critical points for wall
separations $H/\sigma_\text{c}=10$, and 2, where we find a better
agreement between simulation and theory profiles.

\begin{figure}[htbp]
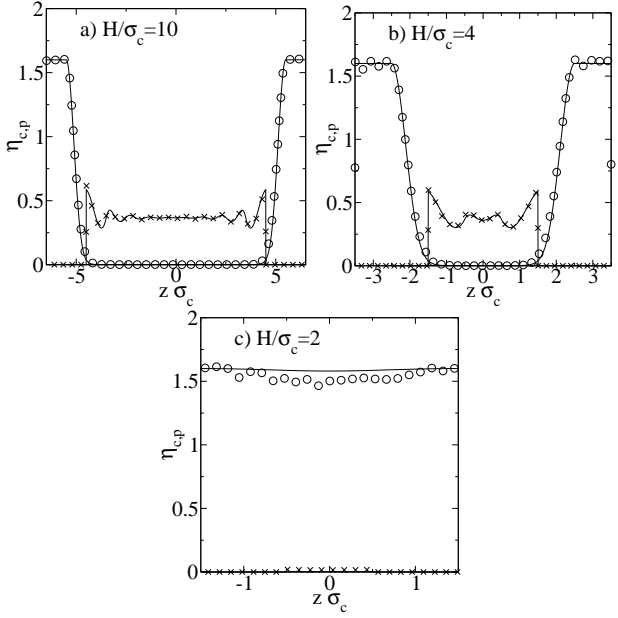

\includegraphics[width=4cm] {fig13a.eps}
\includegraphics[width=4cm] {fig13b.eps}
\includegraphics[width=4cm] {fig13c.eps}
\caption{Density profiles of the AOV model between two parallel
  semi-permeable walls with varying separation distance: a)
  $H_\text{c}/\sigma_\text{c}$=10, b) $H_\text{c}/\sigma_\text{c}$=4,
  and c) $H_\text{c}/\sigma_\text{c}$=2 at statepoint B of the phase
  diagram of Fig.\ \ref{F:stot2}, i.e., for polymer reservoir packing
  fraction $\eta_\text{p}^\text{r}=1.60 $ and colloid fugacity
  $\ln(z_\text{c} \sigma_\text{c}^3)=7.3$. Shown are results from
  simulations for the density profiles of the colloids (crosses) and
  polymers (circles), and results from DFT (full lines).}
\label{F:sprof1}
\end{figure}

We next discuss the structure of the mixture inside the slit with
semi-permeable walls (Fig.\ \ref{F:sprof1}). We fix the fugacities of both species to
those at statepoint B ($\eta_\text{p}^\text{r}$=1.60 and
$\ln(z_\text{c} \sigma_\text{c}^3)$=7.3), see Fig.\ \ref{F:stot2}.
The statepoint B is in the liquid part of the bulk phase diagram
phase. The liquid fills slits with wall separations
$H_\text{c}/\sigma_\text{c}=10$, and 4, while for
$H_\text{c}/\sigma_\text{c}=2$ the slit is filled with gas. This
is an indication of capillary evaporation consistent with the
findings of Sec. \ref{phds}.  The liquid phase is characterized by
structureless polymer profiles, and a layering of colloids for
both $H_\text{c}/\sigma_\text{c}=10$, and 4.  Note that we did not
observe such layering for the statepoints at coexistence (see
previous section).  Simulation and theory are in good agreement at all wall separations.  

\subsection{Two dimensional limit}
\label{2dlim}
We next analyze the dimensional crossover from three to two spatial
dimensions by reducing the distance of the hard walls towards
$H/\sigma_\text{c} \rightarrow 1$.  The two-dimensional system
encountered for $H/\sigma_\text{c} = 1$ is identical to a
two-dimensional mixture of colloidal hard disc and ideal polymer
discs. Two-dimensional mixtures were previously studied with both
theory \cite{JLMR99,MSHL02}, and experiments \cite{TLJL03}.  For $H$
very close to $\sigma_\text{p}$ the polymer reservoir packing fraction scale as

\begin{equation}
  \eta_\text{p}^\text{r} \sim
  \frac{\pi}{6}\frac{N_\text{p} \sigma_\text{p}^3}{A (H-\sigma_\text{p})}.
\end{equation}
We eliminate the divergence using scaled variables for the polymer
reservoir packing fraction
$\eta_\text{p}^\text{r}(H-\sigma_\text{p})/H$ and for the colloidal
fugacity $z_\text{c}(H-\sigma_\text{c})/H$.  Effectively, we map the
three-dimensional system with packing fractions $\eta_i = \pi
\sigma_i^3 N_\text{i}/(6AH)$ to the two-dimensional system with
packing fractions $\eta_i = \pi \sigma_i^2 N_i/(4 A)$, where
$i=\text{c, p}$.

\begin{figure}[htbp]
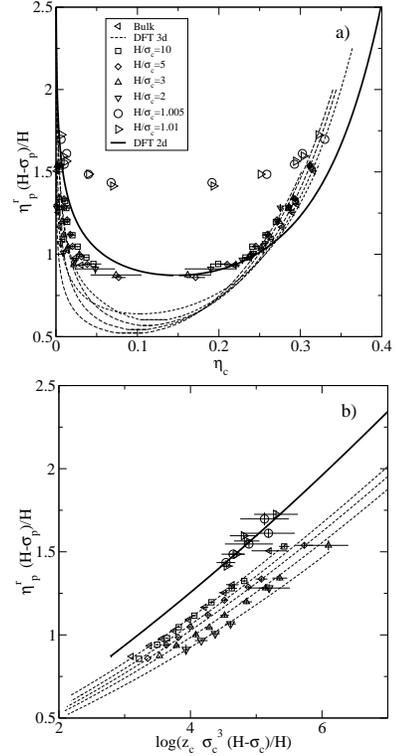

  \begin{center}
    \includegraphics[height=5cm]{fig14a.eps}
    \includegraphics[height=5cm]{fig14b.eps}
    \caption{Phase diagrams showing the binodals for the AOV model
      between parallel hard walls with separation distance
      $H/\sigma_\text{c}=\infty, 10, 5, 3, 2, 1.01,$ and
      $1.005$. Shown are results from simulation (symbols), from
      three-dimensional DFT (dashed curves), and from DFT for the AOV
      model in two dimensions (full lines). a) The gas-liquid binodal
      as a function of the scaled variable $\eta_\text{p}^\text{r}
      (H-\sigma_\text{p})/H$ and $\eta_\text{c}$; b) The gas-liquid
      binodals as a function of the scaled variables $z_\text{c}
      \sigma_\text{c}^3(H-\sigma_\text{c})/H$ and
      $\eta_\text{p}^\text{r} (H-\sigma_\text{p})/H$.}
    \label{F:scal}
  \end{center}
\end{figure}

In Fig.\ \ref{F:scal}(a) we plot the phase diagrams in  the scaled
$\eta_\text{p}^\text{r}-\eta_\text{c}$ representation and we
observe that the binodals for $H/\sigma_\text{c}=1.01$ and
$H/\sigma_\text{c}=1.005$ are superimposed, demonstrating that
this is a reliable estimate for the binodals of the 2-dimensional
system. The comparison with a two-dimensional DFT \cite{MSHL02}
equivalent to a two-dimensional free volume theory \cite{JLMR99}
indicates poorer agreement as in the three-dimensional case. The
discrepancy in the critical polymer reservoir packing fraction is
very substantial. We also find that in this representation the
binodals of the three-dimensional system of the slits collapse
over the bulk binodal, indicating a scaling of the critical value
of $\eta_\text{p}^\text{r}$ as
$(\eta_\text{p}^\text{r})_\text{crit} \sim 1/
(H-\sigma_\text{p})$. In Fig.\ \ref{F:scal}(b) we see that the
collapse of the binodals onto a master curve in  the scaled
$\eta_\text{p}^\text{r}-$ scaled $z_\text{c}$ representation is
not as good as in the other representation, moreover the sequence
of binodals is inverted, indicating that the critical value of the
colloid fugacity scales differently. 

\subsection{Kelvin equation}
\label{S:k2}
In this section, we compare the simulation results with the
predictions of the Kelvin equations that we derived in
section \ref{S:k1}.
First, we address the relationship of the parameter $h$, we used in the Kelvin equations, to our model parameter $H$.
There are two, a priori equivalent,  choices  that we investigate, namely $h=H$, and $h=H-\sigma_c$.
Second, the Kelvin equations need, as a input, the difference of the gas and liquid tensions at the wall interface.
Since this data are not readily available we will assume, in the case of capillary condensation, the relation $\gamma_{wg}-\gamma_{wl}=\gamma_{lg}$, strictly valid only in the complete wetting regime, to hold at all state points considered.
Likewise, for the capillary evaporation case, we assume $\gamma_{wl}-\gamma_{wg}=\gamma_{lg}$, valid in the complete drying regime to hold at all state points. For the liquid-gas interfacial $\gamma_{\rm lg}$ tension we use DFT data from Ref.\ \cite{Brader2000}.

\begin{figure}[htbp]
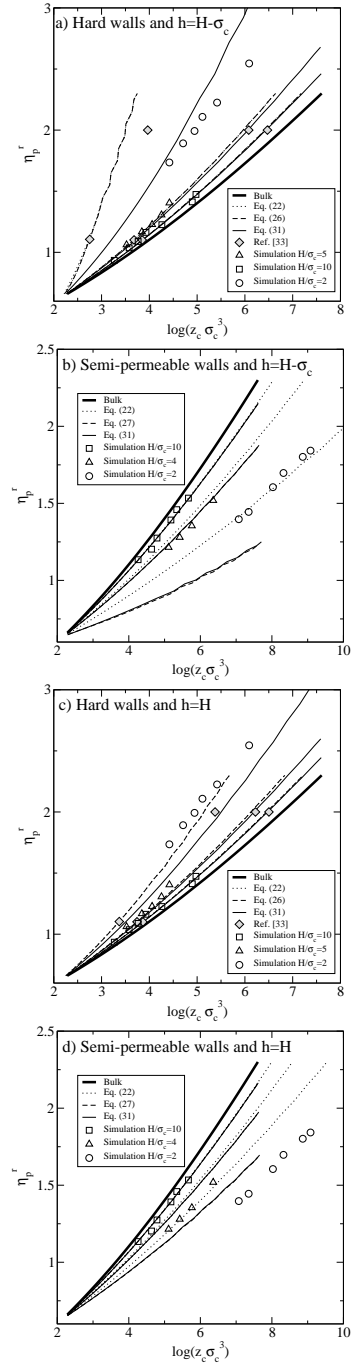

  \begin{center}
    \includegraphics[height=4.5cm]{fig15a.eps}
    \includegraphics[height=4.5cm]{fig15b.eps}
    \includegraphics[height=4.5cm]{fig15c.eps}
    \includegraphics[height=4.5cm]{fig15d.eps}
    \caption{Phase diagram of colloid-polymer mixtures in the  colloidal chemical
      potential $\log(z_\text{c} \sigma^3_\text{c})$ and the polymer reservoir packing fraction $\eta_\text{p}^\text{r}$ representation.
      a) Results for slits with hard walls, and $h=H-\sigma_c$.
      b) Results for the slit with semi-permeable walls , and $h=H-\sigma_c$.
      c) Results for the slits with hard walls, and $h=H$.
      d) Results for the slit with semi-permeable walls, and $h=H$.
      In a) and c) the simulation results (symbols) are compared with the prediction of the Kelvin equations (\ref{EQmupConstant}), (\ref{EQkelvinEMBMcondensationP}), and (\ref{EQkelvinnew}) assuming complete wetting ($\gamma_{wg}-\gamma_{wl}=\gamma_{lg}$). The filled diamonds are results from Eq. (\ref{EQkelvinEMBMcondensationP}) using the difference in wall tensions $\gamma_{wg}-\gamma_{wl}$ taken from Ref.\ \cite{MSAF03}. In b) and d) the simulation results (symbols) are compared with the prediction of the Kelvin equations (\ref{EQmupConstant}), (\ref{EQkelvinEMBMevaporationP}), and (\ref{EQkelvinnew}) assuming complete drying ($\gamma_{wl}-\gamma_{wg}=\gamma_{lg}$). For the liquid-gas interfacial  tension $\gamma_{\rm lg}$  we used  DFT data from Ref.\ \cite{Brader2000}. Few lines are superimposed, see text for the explanation.}
    \label{F:hkelv}
  \end{center}
\end{figure}

Figs. \ref{F:hkelv}(a) and \ref{F:hkelv}(c) display the simulation
results for the hard wall slit together with the predictions of
the  Kelvin equations (\ref{EQmupConstant}),
(\ref{EQkelvinEMBMcondensationP}), and (\ref{EQkelvinnew})  for
$h=H-\sigma_\text{c}$, and $h=H$ respectively. The Kelvin equation
(\ref{EQmupConstant}), derived from  the path with constant
polymer chemical potential, is superimposed at all separation
distances with the Kelvin equation
(\ref{EQkelvinEMBMcondensationP}) derived using the constant
pressure path. This is consistent with the observation of Aarts
and Lekkerkerker \cite{DGALA04}. Now we can offer an alternative
explanation. As shown in Fig \ref{Fig4}, for the capillary
condensation case the bulk in the gas phase (polymer rich phase),
and the path with constant polymer chemical potential is almost
equivalent to the constant pressure path. In other words, the bulk
reference point for Eq. (\ref{EQmupConstant}) and Eq.
(\ref{EQkelvinEMBMcondensationP}) is very similar in the case of
capillary condensation. The Kelvin equation (\ref{EQkelvinnew})
derived using a normal path, predict a smaller shift with respect
to Eq.  (\ref{EQkelvinEMBMcondensationP}). To estimate the error
introduced by the complete wetting approximation
($\gamma_{wg}-\gamma_{wl}=\gamma_{lg}$), we show  few points
(filled diamonds) predicted by Eq.\
(\ref{EQkelvinEMBMcondensationP}) using the actual difference in
wall tensions, as published in our previous work \cite{MSAF03}.

The Figs. \ref{F:hkelv}(b) and \ref{F:hkelv}(d) display the simulation
results for the semi-permeable wall slit together with the
predictions  of the Kelvin equations (\ref{EQmupConstant}),
(\ref{EQkelvinEMBMevaporationP}), and (\ref{EQkelvinnew})  for
$h=H-\sigma_\text{c}$, and $h=H$ respectively. The prediction of
Eq. (\ref{EQkelvinEMBMevaporationP}) and (\ref{EQkelvinnew}) are
superimposed at all  separation distances considered. This is
surprising since they are derived from very different "paths", as
shown in Fig. \ref{Fig4}.

We can conclude that the Kelvin equation we derived using a novel
approach, gives predictions that are consistent, and in
quantitatively agreement with the prediction of the classic
equation. In addition our equation is the same for capillary
condensation and evaporation, and the choice of  reference state,
which is different for both phenomena, is avoided. As the Kelvin
equation is based on macroscopic arguments, is surprising that we
find reasonable quantitative agreement for nearly two-dimensional
systems. The shift of the critical polymer fugacity towards higher
values upon increasing confinement, as found in simulations, is
not reproduced, because the Kelvin equation is entirely based on
properties of the (semi)infinite system. Finally, the two choices
of $h$ we presented give essentially the same results for wall
separations H$/\sigma_c$ as small as 4.

\section{Conclusions}
\label{conc}
In conclusion, we have studied the effect of strong confinement
provided by two parallel walls on the phase behavior and structure of
model mixtures of colloids and polymers of size ratio $q=1$. The
densities of the gas and liquid phases at coexistence, as well as the
chemical potentials were computed by GEMC simulations and DFT.  Two
different models of confining walls were investigated: i) Hard walls,
impenetrable to both colloids and polymers, as a model for glass walls
in contact with colloid-polymer mixtures were found to stabilize the
liquid phase for statepoints that lie in the gas part of the bulk
phase diagram; this effect is referred to as capillary
condensation. ii) Semi-permeable walls, impenetrable to colloids but
penetrable to polymers, could be experimentally realized using polymer
coated substrates \cite{WKW03-a}. If the coating density is not too
high the polymer brushes can act as impenetrable for colloids while
being penetrable for polymers.  We find that the effect of
semi-permeable walls is to stabilize the gas phase for statepoints in
the liquid part of the bulk phase diagram; this effect is referred to
as capillary evaporation.  Both capillary evaporation and condensation
are consistently predicted by GEMC simulations and DFT.  The
differences between simulations and DFT in the vicinity of the
critical point are confirmed for bulk mixtures and were found to be
larger for the confined mixtures. The differences reach a maximum in
the limit of two-dimensional colloid-polymer mixtures.

We have studied the structure of the mixture between parallel
walls by measuring the density profiles in the direction normal to
the confining walls. For the liquid phase, rich in colloids and
poor in polymers, we found layering of colloids with an
oscillation period roughly equal to the diameter of the particles
for all wall separations and statepoints considered.  In the case
of semi-permeable walls, the structure, i.e.\ the layering of
colloids and the adsorption or desorption of gas layers at the
semi-permeable walls, depends strongly on the statepoint and on
the lengthscale of the confinement.  For the gas phase, rich in
polymers and poor in colloids, we found flat polymer profiles with
moderate desorption of polymers from the hard walls.  We found
that density oscillations for colloids and polymers are correlated
in the liquid phase and anti-correlated in the gas phase. This can
be understood by the following argument. In the liquid phase the
fraction of polymers is small and a polymer is always surrounded
by other colloids to which it interacts with an hard-core
potential.  Clearly the polymer structure must be similar to that
of the colloids.  In the gas phase the fraction of polymers is
large with respect to the colloids and the entropy is increased by
segregation of colloids, so if a region is locally denser in
polymer than the average polymer density, it will be more dilute
in colloids.  The comparison between simulation and DFT is overall
good, with small differences in the vicinity of the critical
point.  Our findings of capillary condensation for hard walls and
capillary evaporation for semi-permeable walls are consistent with
the experimental findings of Aarts \et \cite{DGALA04} and Wijting
\et \cite{WKW03,WKW03-a}. Nevertheless the comparison is only
qualitative. Experiments with better controlled geometries are
needed to relate directly to our predictions. The use of the
surface force apparatus, for example, should be possible for
colloid-polymer mixtures. Simulations and theory should proceed to
more realistic models for colloid-polymer mixtures \cite{Bolhuis2002,vink2005a,vink2005c}.
For example, the inclusion of excluded volume interactions between polymer
coils has given accurate results for the bulk phase diagram
\cite{Bolhuis2002}. The use of realistic models for confined
mixtures should help understanding the surface phase behavior of
real colloid-polymer mixtures. Work along this line is in
progress.

\acknowledgments
We thank Bob Evans, Dirk G.\ A.\ L.\ Aarts, and Henk N.\ W.\ Lekkerkerker for many useful and inspiring discussions.
This work is part of the research program of the {\em
Stichting voor Fundamenteel Onderzoek der Materie} (FOM), that is
financially supported by the {\em Nederlandse Organisatie voor
Wetenschappelijk Onderzoek} (NWO).  We thank the Dutch National
Computer Facilities foundation for access to the SGI Origin 3800 and
SGI Altix 3700.  Support by the DFG SFB TR6 ``Physics of colloidal
dispersions in external fields'' is acknowledged.

\bibliographystyle{apsrev}
\bibliography{references}

\begin{thebibliography}{51}
\expandafter\ifx\csname natexlab\endcsname\relax\def\natexlab#1{#1}\fi
\expandafter\ifx\csname bibnamefont\endcsname\relax
  \def\bibnamefont#1{#1}\fi
\expandafter\ifx\csname bibfnamefont\endcsname\relax
  \def\bibfnamefont#1{#1}\fi
\expandafter\ifx\csname citenamefont\endcsname\relax
  \def\citenamefont#1{#1}\fi
\expandafter\ifx\csname url\endcsname\relax
  \def\url#1{\texttt{#1}}\fi
\expandafter\ifx\csname urlprefix\endcsname\relax\def\urlprefix{URL }\fi
\providecommand{\bibinfo}[2]{#2}
\providecommand{\eprint}[2][]{\url{#2}}

\bibitem[{\citenamefont{{J. S. Rowlinson} and {B. Widom}}(2002)}]{JRBW82}
\bibinfo{author}{\bibnamefont{{J. S. Rowlinson}}} \bibnamefont{and}
  \bibinfo{author}{\bibnamefont{{B. Widom}}}, \emph{\bibinfo{title}{Molecular
  {T}heory of {C}apillarity}} (\bibinfo{publisher}{Dover},
  \bibinfo{year}{2002}).

\bibitem[{\citenamefont{Gelb et~al.}(1999)\citenamefont{Gelb, {K. E. Gubbins},
  {R. Radhakrishnan}, and {M. Sliwinska-Bartkowiac}}}]{Gelb1999}
\bibinfo{author}{\bibfnamefont{L.~D.} \bibnamefont{Gelb}},
  \bibinfo{author}{\bibnamefont{{K. E. Gubbins}}},
  \bibinfo{author}{\bibnamefont{{R. Radhakrishnan}}}, \bibnamefont{and}
  \bibinfo{author}{\bibnamefont{{M. Sliwinska-Bartkowiac}}},
  \bibinfo{journal}{Rep. Prog. Phys.} \textbf{\bibinfo{volume}{62}},
  \bibinfo{pages}{1573} (\bibinfo{year}{1999}).

\bibitem[{\citenamefont{{R. Evans}}(1990)}]{RE90}
\bibinfo{author}{\bibnamefont{{R. Evans}}}, \bibinfo{journal}{J. Phys.:
  Condens. Matter} \textbf{\bibinfo{volume}{2}}, \bibinfo{pages}{8989}
  (\bibinfo{year}{1990}).

\bibitem[{\citenamefont{{R. Evans} and {U. Marini Bettolo
  Marconi}}(1987)}]{RE87}
\bibinfo{author}{\bibnamefont{{R. Evans}}} \bibnamefont{and}
  \bibinfo{author}{\bibnamefont{{U. Marini Bettolo Marconi}}},
  \bibinfo{journal}{J. Chem. Phys.} \textbf{\bibinfo{volume}{86}},
  \bibinfo{pages}{7138} (\bibinfo{year}{1987}).

\bibitem[{\citenamefont{Meijer and Frenkel}(1991)}]{EJM91}
\bibinfo{author}{\bibfnamefont{E.~J.} \bibnamefont{Meijer}} \bibnamefont{and}
  \bibinfo{author}{\bibfnamefont{D.}~\bibnamefont{Frenkel}},
  \bibinfo{journal}{Phys. Rev. Lett.} \textbf{\bibinfo{volume}{67}},
  \bibinfo{pages}{1110} (\bibinfo{year}{1991}).

\bibitem[{\citenamefont{Meijer and Frenkel}(1994)}]{EJM94}
\bibinfo{author}{\bibfnamefont{E.~J.} \bibnamefont{Meijer}} \bibnamefont{and}
  \bibinfo{author}{\bibfnamefont{D.}~\bibnamefont{Frenkel}},
  \bibinfo{journal}{J. Chem. Phys.} \textbf{\bibinfo{volume}{1994}},
  \bibinfo{pages}{6873} (\bibinfo{year}{1994}).

\bibitem[{\citenamefont{Ilett et~al.}(1994)\citenamefont{Ilett, Orrock, Poon,
  and Pusey}}]{SMI95}
\bibinfo{author}{\bibfnamefont{S.~M.} \bibnamefont{Ilett}},
  \bibinfo{author}{\bibfnamefont{A.}~\bibnamefont{Orrock}},
  \bibinfo{author}{\bibfnamefont{W.~C.~K.} \bibnamefont{Poon}},
  \bibnamefont{and} \bibinfo{author}{\bibfnamefont{P.~N.} \bibnamefont{Pusey}},
  \bibinfo{journal}{Phys. Rev. E} \textbf{\bibinfo{volume}{51}},
  \bibinfo{pages}{1344} (\bibinfo{year}{1994}).

\bibitem[{\citenamefont{{M. Dijkstra} et~al.}(1999)\citenamefont{{M. Dijkstra},
  {J. M. Brader}, and {R. Evans}}}]{MDJMB99}
\bibinfo{author}{\bibnamefont{{M. Dijkstra}}},
  \bibinfo{author}{\bibnamefont{{J. M. Brader}}}, \bibnamefont{and}
  \bibinfo{author}{\bibnamefont{{R. Evans}}}, \bibinfo{journal}{J. Phys.:
  Condens. Matter} \textbf{\bibinfo{volume}{11}}, \bibinfo{pages}{10079}
  (\bibinfo{year}{1999}).

\bibitem[{\citenamefont{{M. Dijkstra}}(2001)}]{MD01}
\bibinfo{author}{\bibnamefont{{M. Dijkstra}}}, \bibinfo{journal}{Curr. Opinion
  in Coll. \& Interf. Sci} \textbf{\bibinfo{volume}{6}}, \bibinfo{pages}{372}
  (\bibinfo{year}{2001}).

\bibitem[{\citenamefont{{W. C. K. Poon}}(2002)}]{WCKP02}
\bibinfo{author}{\bibnamefont{{W. C. K. Poon}}}, \bibinfo{journal}{J. Phys.:
  Condens. Matter} \textbf{\bibinfo{volume}{14}}, \bibinfo{pages}{R859}
  (\bibinfo{year}{2002}).

\bibitem[{\citenamefont{Tuinier et~al.}(2003)\citenamefont{Tuinier, Rieger, and
  de~Kruif}}]{RT03}
\bibinfo{author}{\bibfnamefont{R.}~\bibnamefont{Tuinier}},
  \bibinfo{author}{\bibfnamefont{J.}~\bibnamefont{Rieger}}, \bibnamefont{and}
  \bibinfo{author}{\bibfnamefont{C.~G.} \bibnamefont{de~Kruif}},
  \bibinfo{journal}{Adv. Coll. Interf. Sci.} \textbf{\bibinfo{volume}{103}},
  \bibinfo{pages}{1} (\bibinfo{year}{2003}).

\bibitem[{\citenamefont{{J. M. Brader} et~al.}(2001)\citenamefont{{J. M.
  Brader}, {M. Dijkstra}, and {R. Evans}}}]{JMB01}
\bibinfo{author}{\bibnamefont{{J. M. Brader}}},
  \bibinfo{author}{\bibnamefont{{M. Dijkstra}}}, \bibnamefont{and}
  \bibinfo{author}{\bibnamefont{{R. Evans}}}, \bibinfo{journal}{Phys. Rev. E}
  \textbf{\bibinfo{volume}{63}}, \bibinfo{pages}{041405}
  (\bibinfo{year}{2001}).

\bibitem[{\citenamefont{{J. M. Brader} et~al.}(2003)\citenamefont{{J. M.
  Brader}, {R. Evans}, and {M. Schmidt}}}]{JMB03}
\bibinfo{author}{\bibnamefont{{J. M. Brader}}},
  \bibinfo{author}{\bibnamefont{{R. Evans}}}, \bibnamefont{and}
  \bibinfo{author}{\bibnamefont{{M. Schmidt}}}, \bibinfo{journal}{Mol. Phys.}
  \textbf{\bibinfo{volume}{101}}, \bibinfo{pages}{3349} (\bibinfo{year}{2003}).

\bibitem[{\citenamefont{{D. G. A. L. Aarts} et~al.}(2004)\citenamefont{{D. G.
  A. L. Aarts}, {R. P. A. Dullens}, {H. N. W. Lekkerker}, {D. Bonn}, and {R.
  van Roij}}}]{DGALA04a}
\bibinfo{author}{\bibnamefont{{D. G. A. L. Aarts}}},
  \bibinfo{author}{\bibnamefont{{R. P. A. Dullens}}},
  \bibinfo{author}{\bibnamefont{{H. N. W. Lekkerker}}},
  \bibinfo{author}{\bibnamefont{{D. Bonn}}}, \bibnamefont{and}
  \bibinfo{author}{\bibnamefont{{R. van Roij}}}, \bibinfo{journal}{J. Chem.
  Phys.} \textbf{\bibinfo{volume}{120}}, \bibinfo{pages}{1973}
  (\bibinfo{year}{2004}).

\bibitem[{\citenamefont{{P. P. F. Wessels}
  et~al.}(2004{\natexlab{a}})\citenamefont{{P. P. F. Wessels}, {M. Schmidt},
  and {H. L\"{o}wen}}}]{PPFW04a}
\bibinfo{author}{\bibnamefont{{P. P. F. Wessels}}},
  \bibinfo{author}{\bibnamefont{{M. Schmidt}}}, \bibnamefont{and}
  \bibinfo{author}{\bibnamefont{{H. L\"{o}wen}}}, \bibinfo{journal}{J. Phys.:
  Condens. Matter} \textbf{\bibinfo{volume}{16}}, \bibinfo{pages}{L1}
  (\bibinfo{year}{2004}{\natexlab{a}}).

\bibitem[{\citenamefont{Fortini et~al.}(2005)\citenamefont{Fortini, Dijkstra,
  Schmidt, and Wessels}}]{Fortini2005a}
\bibinfo{author}{\bibfnamefont{A.}~\bibnamefont{Fortini}},
  \bibinfo{author}{\bibfnamefont{M.}~\bibnamefont{Dijkstra}},
  \bibinfo{author}{\bibfnamefont{M.}~\bibnamefont{Schmidt}}, \bibnamefont{and}
  \bibinfo{author}{\bibfnamefont{P.~P.~F.} \bibnamefont{Wessels}},
  \bibinfo{journal}{Phys. Rev. E} \textbf{\bibinfo{volume}{71}},
  \bibinfo{pages}{051403} (\bibinfo{year}{2005}).

\bibitem[{\citenamefont{{P. P. F. Wessels}
  et~al.}(2004{\natexlab{b}})\citenamefont{{P. P. F. Wessels}, {M. Schmidt},
  and {H. L\"owen}}}]{PPFW04}
\bibinfo{author}{\bibnamefont{{P. P. F. Wessels}}},
  \bibinfo{author}{\bibnamefont{{M. Schmidt}}}, \bibnamefont{and}
  \bibinfo{author}{\bibnamefont{{H. L\"owen}}}, \bibinfo{journal}{J. Phys.:
  Condens. Matter} \textbf{\bibinfo{volume}{16}}, \bibinfo{pages}{S4169}
  (\bibinfo{year}{2004}{\natexlab{b}}).

\bibitem[{\citenamefont{{R. L. C. Vink} and {J.
  Horbach}}(2004{\natexlab{a}})}]{RLCV04c}
\bibinfo{author}{\bibnamefont{{R. L. C. Vink}}} \bibnamefont{and}
  \bibinfo{author}{\bibnamefont{{J. Horbach}}}, \bibinfo{journal}{J. Chem.
  Phys.} \textbf{\bibinfo{volume}{121}}, \bibinfo{pages}{3253}
  (\bibinfo{year}{2004}{\natexlab{a}}).

\bibitem[{\citenamefont{{R. L. C. Vink} and {J.
  Horbach}}(2004{\natexlab{b}})}]{RLCV04}
\bibinfo{author}{\bibnamefont{{R. L. C. Vink}}} \bibnamefont{and}
  \bibinfo{author}{\bibnamefont{{J. Horbach}}}, \bibinfo{journal}{J. Phys.:
  Condens. Matter} \textbf{\bibinfo{volume}{16}}, \bibinfo{pages}{S3807}
  (\bibinfo{year}{2004}{\natexlab{b}}).

\bibitem[{\citenamefont{{M. Dijkstra} and {R. van Roij}}(2002)}]{MDRvR02}
\bibinfo{author}{\bibnamefont{{M. Dijkstra}}} \bibnamefont{and}
  \bibinfo{author}{\bibnamefont{{R. van Roij}}}, \bibinfo{journal}{Phys. Rev.
  Lett.} \textbf{\bibinfo{volume}{89}}, \bibinfo{pages}{208303}
  (\bibinfo{year}{2002}).

\bibitem[{\citenamefont{Dijkstra and van Roij}(2005)}]{MDRvR05}
\bibinfo{author}{\bibfnamefont{M.}~\bibnamefont{Dijkstra}} \bibnamefont{and}
  \bibinfo{author}{\bibfnamefont{R.}~\bibnamefont{van Roij}},
  \bibinfo{journal}{J. Phys.: Condens. Matter} \textbf{\bibinfo{volume}{17}},
  \bibinfo{pages}{S3507} (\bibinfo{year}{2005}).

\bibitem[{\citenamefont{Dijkstra et~al.}(2005)\citenamefont{Dijkstra, van Roij,
  Fortini, and Roth}}]{MDRvRRRAF05}
\bibinfo{author}{\bibfnamefont{M.}~\bibnamefont{Dijkstra}},
  \bibinfo{author}{\bibfnamefont{R.}~\bibnamefont{van Roij}},
  \bibinfo{author}{\bibfnamefont{A.}~\bibnamefont{Fortini}}, \bibnamefont{and}
  \bibinfo{author}{\bibfnamefont{R.}~\bibnamefont{Roth}},
  \bibinfo{journal}{submitted}  (\bibinfo{year}{2005}).

\bibitem[{\citenamefont{{J. M. Brader} et~al.}(2002)\citenamefont{{J. M.
  Brader}, {R. Evans}, {M. Schmidt}, and {H. L\"{o}wen}}}]{JMB02}
\bibinfo{author}{\bibnamefont{{J. M. Brader}}},
  \bibinfo{author}{\bibnamefont{{R. Evans}}}, \bibinfo{author}{\bibnamefont{{M.
  Schmidt}}}, \bibnamefont{and} \bibinfo{author}{\bibnamefont{{H. L\"{o}wen}}},
  \bibinfo{journal}{J. Phys.: Condens. Matter} \textbf{\bibinfo{volume}{14}},
  \bibinfo{pages}{L1} (\bibinfo{year}{2002}).

\bibitem[{\citenamefont{{D. G. A. L. Aarts} et~al.}(2003)\citenamefont{{D. G.
  A. L. Aarts}, {J. H. van der Wiel}, and {H. N. W. Lekkerker}}}]{DGALA03}
\bibinfo{author}{\bibnamefont{{D. G. A. L. Aarts}}},
  \bibinfo{author}{\bibnamefont{{J. H. van der Wiel}}}, \bibnamefont{and}
  \bibinfo{author}{\bibnamefont{{H. N. W. Lekkerker}}}, \bibinfo{journal}{J.
  Phys.: Condens. Matter} \textbf{\bibinfo{volume}{15}}, \bibinfo{pages}{S245}
  (\bibinfo{year}{2003}).

\bibitem[{\citenamefont{{D. G. A. L. Aarts} and {H. N. W.
  Lekkerker}}(2004)}]{DGALA04}
\bibinfo{author}{\bibnamefont{{D. G. A. L. Aarts}}} \bibnamefont{and}
  \bibinfo{author}{\bibnamefont{{H. N. W. Lekkerker}}}, \bibinfo{journal}{J.
  Phys.: Condens. Matter} \textbf{\bibinfo{volume}{16}}, \bibinfo{pages}{S4231}
  (\bibinfo{year}{2004}).

\bibitem[{\citenamefont{{W. K. Wijting}
  et~al.}(2003{\natexlab{a}})\citenamefont{{W. K. Wijting}, {N. A. M.
  Besseling}, and {M. A. Cohen Stuart}}}]{WKW03}
\bibinfo{author}{\bibnamefont{{W. K. Wijting}}},
  \bibinfo{author}{\bibnamefont{{N. A. M. Besseling}}}, \bibnamefont{and}
  \bibinfo{author}{\bibnamefont{{M. A. Cohen Stuart}}}, \bibinfo{journal}{Phys.
  Rev. Lett.} \textbf{\bibinfo{volume}{90}}, \bibinfo{pages}{196101}
  (\bibinfo{year}{2003}{\natexlab{a}}).

\bibitem[{\citenamefont{{W. K. Wijting}
  et~al.}(2003{\natexlab{b}})\citenamefont{{W. K. Wijting}, {N. A. M.
  Besseling}, and {M. A. Cohen Stuart}}}]{WKW03-a}
\bibinfo{author}{\bibnamefont{{W. K. Wijting}}},
  \bibinfo{author}{\bibnamefont{{N. A. M. Besseling}}}, \bibnamefont{and}
  \bibinfo{author}{\bibnamefont{{M. A. Cohen Stuart}}}, \bibinfo{journal}{J.
  Phys. Chem. B} \textbf{\bibinfo{volume}{107}}, \bibinfo{pages}{10565}
  (\bibinfo{year}{2003}{\natexlab{b}}).

\bibitem[{\citenamefont{Jenkins and Snowden}(1996)}]{PJ96}
\bibinfo{author}{\bibfnamefont{P.}~\bibnamefont{Jenkins}} \bibnamefont{and}
  \bibinfo{author}{\bibfnamefont{M.}~\bibnamefont{Snowden}},
  \bibinfo{journal}{Adv. Coll. Interf. Sci.} \textbf{\bibinfo{volume}{68}},
  \bibinfo{pages}{57} (\bibinfo{year}{1996}).

\bibitem[{\citenamefont{Wessels et~al.}(2005)\citenamefont{Wessels, Schmidt,
  and L\"owen}}]{Wessels2005}
\bibinfo{author}{\bibfnamefont{P.~P.~F.} \bibnamefont{Wessels}},
  \bibinfo{author}{\bibfnamefont{M.}~\bibnamefont{Schmidt}}, \bibnamefont{and}
  \bibinfo{author}{\bibfnamefont{H.}~\bibnamefont{L\"owen}},
  \bibinfo{journal}{Phys. Rev. Lett.} \textbf{\bibinfo{volume}{94}},
  \bibinfo{pages}{078303} (\bibinfo{year}{2005}).

\bibitem[{\citenamefont{{M. Schmidt}
  et~al.}(2002{\natexlab{a}})\citenamefont{{M. Schmidt}, {E.
  Sch\"{o}ll-Paschinger}, {J\"{u}rgen K\"{o}finger}, and {Gerard
  Kahl}}}]{MSES03}
\bibinfo{author}{\bibnamefont{{M. Schmidt}}}, \bibinfo{author}{\bibnamefont{{E.
  Sch\"{o}ll-Paschinger}}}, \bibinfo{author}{\bibnamefont{{J\"{u}rgen
  K\"{o}finger}}}, \bibnamefont{and} \bibinfo{author}{\bibnamefont{{Gerard
  Kahl}}}, \bibinfo{journal}{J. Phys.: Condens. Matter}
  \textbf{\bibinfo{volume}{14}}, \bibinfo{pages}{12099}
  (\bibinfo{year}{2002}{\natexlab{a}}).

\bibitem[{\citenamefont{Wessels et~al.}(2003)\citenamefont{Wessels, Schmidt,
  and L\"owen}}]{PPFW03}
\bibinfo{author}{\bibfnamefont{P.~P.~F.} \bibnamefont{Wessels}},
  \bibinfo{author}{\bibfnamefont{M.}~\bibnamefont{Schmidt}}, \bibnamefont{and}
  \bibinfo{author}{\bibfnamefont{H.}~\bibnamefont{L\"owen}},
  \bibinfo{journal}{Phys. Rev. E} \textbf{\bibinfo{volume}{68}},
  \bibinfo{pages}{061404} (\bibinfo{year}{2003}).

\bibitem[{\citenamefont{{I. O. G\"otze} et~al.}(2003)\citenamefont{{I. O.
  G\"otze}, {J. M. Brader}, {M. Schmidt}, and {H. L\"owen}}}]{IOG+03}
\bibinfo{author}{\bibnamefont{{I. O. G\"otze}}},
  \bibinfo{author}{\bibnamefont{{J. M. Brader}}},
  \bibinfo{author}{\bibnamefont{{M. Schmidt}}}, \bibnamefont{and}
  \bibinfo{author}{\bibnamefont{{H. L\"owen}}}, \bibinfo{journal}{Mol. Phys.}
  \textbf{\bibinfo{volume}{101}}, \bibinfo{pages}{1651} (\bibinfo{year}{2003}).

\bibitem[{\citenamefont{{M. Schmidt} et~al.}(2003)\citenamefont{{M. Schmidt},
  {A. Fortini}, and {M. Dijkstra}}}]{MSAF03}
\bibinfo{author}{\bibnamefont{{M. Schmidt}}}, \bibinfo{author}{\bibnamefont{{A.
  Fortini}}}, \bibnamefont{and} \bibinfo{author}{\bibnamefont{{M. Dijkstra}}},
  \bibinfo{journal}{J. Phys.: Condens. Matter} \textbf{\bibinfo{volume}{15}},
  \bibinfo{pages}{3411} (\bibinfo{year}{2003}).

\bibitem[{\citenamefont{{M. Schmidt} et~al.}(2004)\citenamefont{{M. Schmidt},
  {A. Fortini}, and {M. Dijkstra}}}]{MSAF04}
\bibinfo{author}{\bibnamefont{{M. Schmidt}}}, \bibinfo{author}{\bibnamefont{{A.
  Fortini}}}, \bibnamefont{and} \bibinfo{author}{\bibnamefont{{M. Dijkstra}}},
  \bibinfo{journal}{J. Phys.: Condens. Matter} \textbf{\bibinfo{volume}{16}},
  \bibinfo{pages}{S4159} (\bibinfo{year}{2004}).

\bibitem[{\citenamefont{{S. Asakura} and {F. Oosawa}}(1954)}]{SAFO54}
\bibinfo{author}{\bibnamefont{{S. Asakura}}} \bibnamefont{and}
  \bibinfo{author}{\bibnamefont{{F. Oosawa}}}, \bibinfo{journal}{J. Chem.
  Phys.} \textbf{\bibinfo{volume}{22}}, \bibinfo{pages}{1255}
  (\bibinfo{year}{1954}).

\bibitem[{\citenamefont{Asakura and Oosawa}(1958)}]{Asakura1958}
\bibinfo{author}{\bibfnamefont{S.}~\bibnamefont{Asakura}} \bibnamefont{and}
  \bibinfo{author}{\bibfnamefont{F.}~\bibnamefont{Oosawa}},
  \bibinfo{journal}{J. Polym. Sci.} \textbf{\bibinfo{volume}{33}},
  \bibinfo{pages}{183} (\bibinfo{year}{1958}).

\bibitem[{\citenamefont{{A. Vrij}}(1976)}]{AV76}
\bibinfo{author}{\bibnamefont{{A. Vrij}}}, \bibinfo{journal}{Pure Appl. Chem.}
  \textbf{\bibinfo{volume}{48}}, \bibinfo{pages}{471} (\bibinfo{year}{1976}).

\bibitem[{\citenamefont{{A. Z. Panagiotopoulos}}(1987{\natexlab{a}})}]{AZP87}
\bibinfo{author}{\bibnamefont{{A. Z. Panagiotopoulos}}}, \bibinfo{journal}{Mol.
  Phys.} \textbf{\bibinfo{volume}{61}}, \bibinfo{pages}{813}
  (\bibinfo{year}{1987}{\natexlab{a}}).

\bibitem[{\citenamefont{{A. Z. Panagiotopoulos}}(1987{\natexlab{b}})}]{AZP87b}
\bibinfo{author}{\bibnamefont{{A. Z. Panagiotopoulos}}}, \bibinfo{journal}{Mol.
  Phys.} \textbf{\bibinfo{volume}{62}}, \bibinfo{pages}{701}
  (\bibinfo{year}{1987}{\natexlab{b}}).

\bibitem[{\citenamefont{{D. Frenkel} and {B. Smit}}(2002)}]{DFBS96}
\bibinfo{author}{\bibnamefont{{D. Frenkel}}} \bibnamefont{and}
  \bibinfo{author}{\bibnamefont{{B. Smit}}},
  \emph{\bibinfo{title}{Understanding {M}olecular {S}imulation 2nd edition}},
  vol.~\bibinfo{volume}{1} of \emph{\bibinfo{series}{Computational Science
  Series}} (\bibinfo{publisher}{Academic Press}, \bibinfo{year}{2002}).

\bibitem[{\citenamefont{{B. Smit} and {D. Frenkel}}(1989)}]{BSDF89}
\bibinfo{author}{\bibnamefont{{B. Smit}}} \bibnamefont{and}
  \bibinfo{author}{\bibnamefont{{D. Frenkel}}}, \bibinfo{journal}{Mol. Phys.}
  \textbf{\bibinfo{volume}{68}}, \bibinfo{pages}{951} (\bibinfo{year}{1989}).

\bibitem[{\citenamefont{Schmidt et~al.}(2000)\citenamefont{Schmidt, {H.
  L\"owen}, {J. M. Brader}, and {R. Evans}}}]{Schmidt2000}
\bibinfo{author}{\bibfnamefont{M.}~\bibnamefont{Schmidt}},
  \bibinfo{author}{\bibnamefont{{H. L\"owen}}},
  \bibinfo{author}{\bibnamefont{{J. M. Brader}}}, \bibnamefont{and}
  \bibinfo{author}{\bibnamefont{{R. Evans}}}, \bibinfo{journal}{Phys. Rev.
  Lett.} \textbf{\bibinfo{volume}{85}}, \bibinfo{pages}{1934}
  (\bibinfo{year}{2000}).

\bibitem[{\citenamefont{{H. N. W. Lekkerkerker} et~al.}(1992)\citenamefont{{H.
  N. W. Lekkerkerker}, {W. C. K. Poon}, {P. N. Pusey}, {A. Stroobants}, and
  {P.B. Warren}}}]{HNWL+92}
\bibinfo{author}{\bibnamefont{{H. N. W. Lekkerkerker}}},
  \bibinfo{author}{\bibnamefont{{W. C. K. Poon}}},
  \bibinfo{author}{\bibnamefont{{P. N. Pusey}}},
  \bibinfo{author}{\bibnamefont{{A. Stroobants}}}, \bibnamefont{and}
  \bibinfo{author}{\bibnamefont{{P.B. Warren}}}, \bibinfo{journal}{Europhys.
  Lett.} \textbf{\bibinfo{volume}{20}}, \bibinfo{pages}{559}
  (\bibinfo{year}{1992}).

\bibitem[{\citenamefont{Vink et~al.}(2006)\citenamefont{Vink, Binder, and
  Horbach}}]{vink2005}
\bibinfo{author}{\bibfnamefont{R.~L.~C.} \bibnamefont{Vink}},
  \bibinfo{author}{\bibfnamefont{K.}~\bibnamefont{Binder}}, \bibnamefont{and}
  \bibinfo{author}{\bibfnamefont{J.}~\bibnamefont{Horbach}},
  \bibinfo{journal}{cond-matt/0602348}  (\bibinfo{year}{2006}).

\bibitem[{\citenamefont{{J.-T. Lee} and {M. Robert}}(1999)}]{JLMR99}
\bibinfo{author}{\bibnamefont{{J.-T. Lee}}} \bibnamefont{and}
  \bibinfo{author}{\bibnamefont{{M. Robert}}}, \bibinfo{journal}{Phys. Rev. E}
  \textbf{\bibinfo{volume}{60}}, \bibinfo{pages}{7198} (\bibinfo{year}{1999}).

\bibitem[{\citenamefont{{M. Schmidt}
  et~al.}(2002{\natexlab{b}})\citenamefont{{M. Schmidt}, {H. L\"owen}, {J. M.
  Brader}, and {R. Evans}}}]{MSHL02}
\bibinfo{author}{\bibnamefont{{M. Schmidt}}}, \bibinfo{author}{\bibnamefont{{H.
  L\"owen}}}, \bibinfo{author}{\bibnamefont{{J. M. Brader}}}, \bibnamefont{and}
  \bibinfo{author}{\bibnamefont{{R. Evans}}}, \bibinfo{journal}{J. Phys.:
  Condens. Matter} \textbf{\bibinfo{volume}{14}}, \bibinfo{pages}{9353}
  (\bibinfo{year}{2002}{\natexlab{b}}).

\bibitem[{\citenamefont{{T.-C Lee} et~al.}(2003)\citenamefont{{T.-C Lee},
  {J.-T. Lee}, {D.R. Pilaski}, and {M. Robert}}}]{TLJL03}
\bibinfo{author}{\bibnamefont{{T.-C Lee}}},
  \bibinfo{author}{\bibnamefont{{J.-T. Lee}}},
  \bibinfo{author}{\bibnamefont{{D.R. Pilaski}}}, \bibnamefont{and}
  \bibinfo{author}{\bibnamefont{{M. Robert}}}, \bibinfo{journal}{Physica A}
  \textbf{\bibinfo{volume}{329}}, \bibinfo{pages}{411} (\bibinfo{year}{2003}).

\bibitem[{\citenamefont{Brader and Evans}(2000)}]{Brader2000}
\bibinfo{author}{\bibfnamefont{J.~M.} \bibnamefont{Brader}} \bibnamefont{and}
  \bibinfo{author}{\bibfnamefont{R.}~\bibnamefont{Evans}},
  \bibinfo{journal}{Europhys. Lett.} \textbf{\bibinfo{volume}{49}},
  \bibinfo{pages}{678} (\bibinfo{year}{2000}).

\bibitem[{\citenamefont{Bolhuis et~al.}(2002)\citenamefont{Bolhuis, Louis, and
  Hansen}}]{Bolhuis2002}
\bibinfo{author}{\bibfnamefont{P.~G.} \bibnamefont{Bolhuis}},
  \bibinfo{author}{\bibfnamefont{A.~A.} \bibnamefont{Louis}}, \bibnamefont{and}
  \bibinfo{author}{\bibfnamefont{J.-P.} \bibnamefont{Hansen}},
  \bibinfo{journal}{Phys. Rev. Lett.} \textbf{\bibinfo{volume}{89}},
  \bibinfo{pages}{128302} (\bibinfo{year}{2002}).

\bibitem[{\citenamefont{Vink and Schmidt}(2005)}]{vink2005a}
\bibinfo{author}{\bibfnamefont{R.~L.~C.} \bibnamefont{Vink}} \bibnamefont{and}
  \bibinfo{author}{\bibfnamefont{M.}~\bibnamefont{Schmidt}},
  \bibinfo{journal}{Phys. Rev. E} \textbf{\bibinfo{volume}{71}},
  \bibinfo{pages}{051406} (\bibinfo{year}{2005}).

\bibitem[{\citenamefont{Vink et~al.}(2005)\citenamefont{Vink, Jusufi,
  Dzubiella, and Likos}}]{vink2005c}
\bibinfo{author}{\bibfnamefont{R.~L.~C.} \bibnamefont{Vink}},
  \bibinfo{author}{\bibfnamefont{A.}~\bibnamefont{Jusufi}},
  \bibinfo{author}{\bibfnamefont{J.}~\bibnamefont{Dzubiella}},
  \bibnamefont{and} \bibinfo{author}{\bibfnamefont{C.~N.} \bibnamefont{Likos}},
  \bibinfo{journal}{Physical Review E} \textbf{\bibinfo{volume}{72}},
  \bibinfo{pages}{030401} (\bibinfo{year}{2005}).

\end{thebibliography}

\end{document}